\documentclass[a4paper,10pt,fleqn]{article}
\usepackage{epsfig}
\newcommand{\anl}{\\[1ex]}
\newcommand{\bsg}{$b \rightarrow s \gamma$}
\def\lsim{\mathrel{\rlap{\lower4pt\hbox{\hskip1pt$\sim$}}
    \raise1pt\hbox{$<$}}}         
\def\gsim{\mathrel{\rlap{\lower4pt\hbox{\hskip1pt$\sim$}}
    \raise1pt\hbox{$>$}}}         
\def\esim{\mathrel{\rlap{\raise2pt\hbox{$\sim$}}
    \lower1pt\hbox{$-$}}}         

\begin{document}

\rightline{UUITP-11/97}
\rightline{MPI-PhT/97-27}
\rightline{April 1997}

\bigskip
\bigskip

\centerline{\Large \bf Neutralino Relic Density including 
Coannihilations}

\bigskip
\bigskip

\centerline{\large Joakim Edsj{\"o}\footnote{E-mail address: 
edsjo@teorfys.uu.se}}
\smallskip
\centerline{\em Department of Theoretical Physics, Uppsala University}
\centerline{\em Box 803, SE-751 08 Uppsala, Sweden}

\bigskip

\centerline{\large Paolo Gondolo\footnote{E-mail address: 
gondolo@mppmu.mpg.de}}
\smallskip
\centerline{\em Max-Planck-Institut f\"ur Physik (Werner-Heisenberg-Institut)}
\centerline{\em F\"ohringer Ring 6, 80805 M\"unchen, Germany}

\bigskip


\bigskip

\begin{abstract}
  
  We evaluate the relic density of the lightest neutralino, the
  lightest supersymmetric particle, in the Minimal Supersymmetric
  extension of the Standard Model (MSSM).  For the first time, we
  include all coannihilation processes between neutralinos and
  charginos for any neutralino mass and composition.  We use the most
  sophisticated routines for integrating the cross sections and the
  Boltzmann equation. We properly treat (sub)threshold and resonant
  annihilations. We also include one-loop corrections to neutralino
  masses.  We find that coannihilation processes are important not
  only for light higgsino-like neutralinos, as pointed out before, but
  also for heavy higgsinos and for mixed and gaugino-like neutralinos.
  Indeed, coannihilations should be included whenever $|\mu| \lsim 2
  |M_1|$, independently of the neutralino composition. When $|\mu|
  \sim |M_1|$, coannihilations can increase or decrease the relic
  density in and out of the cosmologically interesting region. We find
  that there is still a window of light higgsino-like neutralinos that
  are viable dark matter candidates and that coannihilations shift the
  cosmological upper bound on the neutralino mass from 3 to 7 TeV.

\end{abstract}



\section{Introduction}

In the near future, it may become possible to constrain supersymmetry
from high precision measurements of the cosmological parameters
\cite{CMBdetexp,CMBdettheory}, among which the dark matter density.
It is therefore of great importance to calculate the relic density of
the lightest neutralino as accurately as possible.

The lightest neutralino is one of the most promising candidates for
the dark matter in the Universe. It is believed to be the lightest
stable supersymmetric particle in the Minimal Supersymmetric
extension of the Standard Model (MSSM). It is a linear combination
of the superpartners of the neutral gauge and Higgs bosons.

The relic density of neutralinos in the MSSM has been calculated by
several authors during the years
\cite{relcalc,SWO,GriestSeckel,McDonald,MizutaYamaguchi,DreesNojiri,NeuLoop1}
with various degrees of precision. A complete and precise
calculation including relativistic Boltzmann averaging, subthreshold
and resonant annihilations, and coannihilation processes is the
purpose of this paper.

As pointed out by Griest and Seckel \cite{GriestSeckel} one has to
include coannihilations between the lightest neutralino and other
supersymmetric particles heavier than the neutralino if they are
close in mass.  They considered coannihilations between the lightest
neutralino and the squarks, which occur only accidentally when the
squarks are only slightly heavier than the lightest neutralino. In
contrast, Mizuta and Yamaguchi \cite{MizutaYamaguchi} pointed out an
unavoidable mass degeneracy that greatly affects the neutralino
relic density: the degeneracy between the lightest and
next-to-lightest neutralinos and the lightest chargino when the
neutralino is higgsino-like.  They considered coannihilations
between the lightest neutralino and the lightest chargino, but only
for neutralinos lighter then the $W$ boson and only with an
approximate relic density calculation. Moreover, they did not
consider Higgs bosons in the final states.

Drees and Nojiri \cite{DreesNojiri} included coannihilations into
their relic density calculation, but only between the lightest and
next-to-lightest neutralinos. These coannihilations are not as
important as those studied by Mizuta and Yamaguchi. Recently, Drees et
al.\ \cite{NeuLoop1} re-investigated the relic density of light
higgsino-like neutralinos. They included coannihilations between the
lightest and next-to-lightest neutralinos as well as those between the
lightest neutralino and the lightest chargino.  They do however only
consider $f \bar{f}$, $f \bar{f'}$ and $\gamma W^+$ final states
through $Z$ and $W$ exchange respectively, and do not consider $t$-
and $u$-channel annihilation or Higgs bosons in the final states.

In this paper we perform a full calculation of the neutralino relic
density for any neutralino mass and composition, including for the
first time all coannihilations between neutralinos and charginos. We
properly compute the thermal average, particularly in presence of
thresholds and resonances in the annihilation cross sections. We
include all two-body final states of neutralino-neutralino,
neutralino-chargino and chargino-chargino annihilations.  We leave
coannihilations with squarks \cite{GriestSeckel} for future work,
since they only occur accidentally when the squarks happen to be
close in mass to the lightest neutralino as opposed to the
unavoidable mass degeneracy of the lightest two neutralinos and the
lightest chargino for higgsino-like neutralinos.

In section~\ref{sec:MSSMdef}, we define the MSSM model we use and in
section~\ref{sec:Boltzmann} we describe how we generalize the Gondolo
and Gelmini \cite{GondoloGelmini} formulas to solve the Boltzmann
equation and perform the thermal averages when coannihilations are
included. This is done in a very convenient way by introducing an
effective invariant annihilation rate $W_{\rm eff}$. In
section~\ref{sec:AnnCross} we describe how we calculate all
annihilation cross sections, and in section~\ref{sec:NumMethods} we
outline the numerical methods we use. We then discuss our survey of
supersymmetric models in section~\ref{sec:SelMod}, together with the
experimental constraints we apply.  We finally present our results on
the neutralino relic density in section~\ref{sec:Results} and give
some concluding remarks in section~\ref{sec:Concl}.


\section{Definition of the Supersymmetric model}
\label{sec:MSSMdef}

We work in the framework of the minimal supersymmetric extension of
the standard model defined by, besides the particle content and gauge
couplings required by supersymmetry, the superpotential (the notation
used is similar to Ref.~\cite{haberkane})
\begin{eqnarray}
  W = \epsilon_{ij} \left(
  - {\bf \hat{e}}_{R}^{*} {\bf Y}_E {\bf \hat{l}}^i_{L} {\hat H}^j_1 
  - {\bf \hat{d}}_{R}^{*} {\bf Y}_D {\bf \hat{q}}^i_{L} {\hat H}^j_1 
  + {\bf \hat{u}}_{R}^{*} {\bf Y}_U {\bf \hat{q}}^i_{L} {\hat H}^j_2 
  - \mu {\hat H}^i_1 {\hat H}^j_2 
  \right) 
  \label{superpotential}
\end{eqnarray}
and the soft supersymmetry-breaking potential
\begin{eqnarray}
  \label{Vsoft}
  \lefteqn{V_{{\rm soft}} =  
  \epsilon_{ij} \left(
   {\bf \tilde{e}}_{R}^{*} {\bf A}_E {\bf Y}_E {\bf \tilde{l}}^i_{L} 
H^j_1 
  + {\bf \tilde{d}}_{R}^{*} {\bf A}_D {\bf Y}_D {\bf \tilde{q}}^i_{L} 
H^j_1 
  - {\bf \tilde{u}}_{R}^{*} {\bf A}_U {\bf Y}_U {\bf \tilde{q}}^i_{L} 
H^j_2 
  - B \mu H^i_1 H^j_2 + {\rm h.c.} 
  \right)} \nonumber \\ &&
  + H^{i*}_1 m_1^2 H^i_1 + H^{i*}_2 m_2^2 H^i_2
  \nonumber \\ && +
  {\bf \tilde{q}}_{L}^{i*} {\bf M}_{Q}^{2} {\bf \tilde{q}}^i_{L} + 
  {\bf \tilde{l}}_{L}^{i*} {\bf M}_{L}^{2} {\bf \tilde{l}}^i_{L} + 
  {\bf \tilde{u}}_{R}^{*} {\bf M}_{U}^{2} {\bf \tilde{u}}_{R} + 
  {\bf \tilde{d}}_{R}^{*} {\bf M}_{D}^{2} {\bf \tilde{d}}_{R} + 
  {\bf \tilde{e}}_{R}^{*} {\bf M}_{E}^{2} {\bf \tilde{e}}_{R} 
  \nonumber \\ && +
  {1\over2} M_1 \tilde{B} \tilde{B} +
  {1\over2} M_2 \left( \tilde{W}^3 \tilde{W}^3 +
        2 \tilde{W}^+ \tilde{W}^- \right) +
  {1\over2} M_3 \tilde{g} \tilde{g} . 
\end{eqnarray}
Here $i$ and $j$ are SU(2) indices ($\epsilon_{12} = +1$). The Yukawa
couplings ${\bf Y}$, the soft trilinear couplings ${\bf A}$ and the
soft sfermion masses ${\bf M}$ are $3\times3$ matrices in generation
space. $\bf \hat{e}$, $\bf \hat{l}$, $\bf \hat{u}$, $\bf \hat{d}$ and
$\bf \hat{q}$ are the superfields of the leptons and sleptons and of
the quarks and squarks. A tilde indicates their respective scalar
components.  The $L$ and $R$ subscripts on the sfermion fields refer
to the chirality of their fermionic superpartners. $\tilde{B}$,
$\tilde{W}^3$ and $\tilde{W}^\pm$ are the fermionic superpartners of
the SU(2) gauge fields and $\tilde{g}$ is the gluino field. $\mu$ is
the higgsino mass parameter, $M_1$, $M_2$ and $M_3$ are the gaugino
mass parameters, $B$ is a soft bilinear coupling, and $m^2_{1,2}$ are
Higgs mass parameters.

For $M_1$ and $M_2$ we make the usual GUT assumptions
\begin{eqnarray} \label{eq:M1}
  M_1 & = & \frac{5}{3} M_2 \tan^2 \theta_W \simeq 0.5 M_2 \\
  M_2 & = & \frac{\alpha_{\rm ew}}{\sin^2 \theta_W \alpha_s} M_3
  \simeq 0.3 M_3 \label{eq:M2}
\end{eqnarray}
where $\alpha_{\rm ew}$ is the fine-structure constant and 
$\alpha_{s}$ is the strong coupling constant. 

Electroweak symmetry breaking is caused by both $H^1_1$ and $H^2_2$
acquiring vacuum expectation values,
\begin{equation}
  \langle H^1_1\rangle = v_1 , \qquad \langle H^2_2\rangle = v_2,
\end{equation}
with $g^2(v_1^2+v_2^2) = 2 m_W^2$, with the further assumption that
vacuum expectation values of all other scalar fields (in particular,
squarks and sleptons) vanish. This avoids color and/or charge
breaking vacua.  It is convenient to use expressions for the $Z$
boson mass, $m_Z^2 = {1\over2} (g^2+g'^2) (v_1^2+v_2^2)$ and the
ratio of vacuum expectation values $\tan\beta = v_2/v_1$. $g$ and
$g'$ are the usual SU(2) and U(1) gauge coupling constants.

When diagonalizing the mass matrix for the scalar Higgs fields,
besides a charged and a neutral would-be Goldstone bosons which become
the longitudinal polarizations of the $W^\pm$ and $Z$ gauge bosons,
one finds a neutral CP-odd Higgs boson $A$, two neutral CP-even Higgs
bosons $H_{1,2}$ and a charged Higgs boson $H^{\pm}$. 
Choosing as independent
parameter the mass $m_A$ of the CP-odd Higgs boson, the masses of the
other Higgs bosons are given by
\begin{eqnarray}
  {\cal M}^2_{H} = 
  \left( \matrix{ 
  {m_A^2 \cos^2\beta + m_Z^2 \sin^2\beta + \Delta {\cal M}^2_{11} } &
  {-\sin\beta\cos\beta(m_A^2+m_Z^2) + \Delta {\cal M}^2_{12} }
  \cr
  {-\sin\beta\cos\beta(m_A^2+m_Z^2) + \Delta {\cal M}^2_{21} } &
  {m_A^2 \sin^2\beta + m_Z^2 \cos^2\beta + \Delta {\cal M}^2_{22} }
  } \right)
\end{eqnarray}
\begin{eqnarray}
  m^2_{H^{\pm}} = m_{A}^2+m_W^2 + \Delta_{\pm}.
  \label{mhpm}
\end{eqnarray}
The quantities $\Delta{\cal M}^2_{ij}$ and $\Delta_{\pm}$ are the
leading log two-loop radiative corrections coming from virtual (s)top and
(s)bottom loops, calculated within the effective potential approach
given in \cite{carena} (other references on radiative corrections are
\cite{effpot}).  Diagonalization of $ {\cal M}^2_{H} $ gives the two
CP-even Higgs boson masses, $ m_{H_{1,2}} $, and their mixing angle
$\alpha$ ($ -\pi/2 < \alpha < 0$).
  
The neutralinos $\tilde{\chi}^0_i$ are linear combinations of the 
superpartners of the neutral gauge bosons ${\tilde B}$, $\tilde{W}_3$ 
and of the neutral higgsinos ${\tilde H_1^0}$, ${\tilde H_2^0}$.  In 
this basis, their mass matrix is given by
\begin{eqnarray} \label{eq:neumass}
  {\cal M}_{\tilde \chi^0_{1,2,3,4}} = 
  \left( \matrix{
  {M_1} & 0 & -{g'v_1\over\sqrt{2}} & +{g'v_2\over\sqrt{2}} \cr
  0 & {M_2} & +{gv_1\over\sqrt{2}} & -{gv_2\over\sqrt{2}} \cr
  -{g'v_1\over\sqrt{2}} & +{gv_1\over\sqrt{2}} & \delta_{33} & -\mu 
\cr
  +{g'v_2\over\sqrt{2}} & -{gv_2\over\sqrt{2}} & -\mu & \delta_{44} 
\cr
  } \right)
\end{eqnarray}
where $\delta_{33}$ and $\delta_{44}$ are the most important 
one-loop corrections. These can change the neutralino masses by a few
GeV up or down and are only important when there is a severe mass
degeneracy of the lightest neutralinos and/or charginos.
The expressions for $\delta_{33}$ and $\delta_{44}$ are 
\cite{NeuLoop1,NeuLoop2}
\begin{eqnarray}
  \delta_{33} & = & - \frac{3}{16\pi^2}Y_{b}^2 m_b \sin (2 
  \theta_{\tilde{b}}) {\rm Re} \left[
  B_{0}(Q,b,\tilde{b}_{1})-B_{0}(Q,b,\tilde{b}_{2}) \right] \\
  \delta_{44} & = & - \frac{3}{16\pi^2}Y_{t}^2 m_t \sin (2 
  \theta_{\tilde{t}}) {\rm Re} \left[
  B_{0}(Q,t,\tilde{t}_{1})-B_{0}(Q,t,\tilde{t}_{2}) \right]
\end{eqnarray}
where $m_{b}$ and $m_{t}$ are the masses of the $b$ and $t$ quarks,
$Y_{b}=gm_{b}/\sqrt{2}m_{W}\cos \beta$ and $Y_{t}=gm_{t}/\sqrt{2}m_{W}
\sin \beta$ are the Yukawa couplings of the $b$ and $t$ quark,
$\theta_{\tilde{b}}$ and $\theta_{\tilde{t}}$ are the mixing angles of
the squark mass eigenstates ($\tilde{q}_{1} = \tilde{q}_{L} \cos
\theta_{\tilde{q}} + \tilde{q}_{R} \sin \theta_{\tilde{q}}$) and
$B_{0}$ is the two-point function for which we use the convention in
Ref.~\cite{NeuLoop1,NeuLoop2}.  Expressions for $B_{0}$ can be found in
e.g.\ Ref.~\cite{B-functions}.  For the momentum scale $Q$ we use
$|\mu|$ as suggested in Ref.~\cite{NeuLoop1}.  Note that the loop
corrections depend on the mixing angles of the squarks which in turn
depend on the soft supersymmetry breaking parameters ${\bf A}_{U}$ 
and ${\bf A}_{D}$ in Eq.~(\ref{Vsoft}) (or the parameters $A_{b}$ 
and $A_{t}$ given below).

The neutralino mass matrix, Eq.~(\ref{eq:neumass}), can be 
diagonalized analytically to give four neutral Majorana states,
\begin{equation}
  \tilde{\chi}^0_i = 
  N_{i1} \tilde{B} + N_{i2} \tilde{W}^3 + 
  N_{i3} \tilde{H}^0_1 + N_{i4} \tilde{H}^0_2 ,
\end{equation}
the lightest of which, to be called $\chi$, is then the candidate for
the particle making up the dark matter in the Universe. The gaugino
fraction, $Z_g^i$ of neutralino $i$ is then defined as
\begin{equation}
  Z_g^i = |N_{i1}|^2 + |N_{i2}|^2
\end{equation}
We will call the neutralino higgsino-like if $Z_g<0.01$, mixed if
$0.01 \leq Z_g \leq 0.99$ and gaugino-like if $Z_g>0.99$, where $Z_g
\equiv Z_g^1$ is the gaugino fraction of the lightest neutralino.
Note that the boundaries for what we call gaugino-like and
higgsino-like are somewhat arbitrary and may differ from other
authors.

The charginos are linear combinations of the charged gauge bosons
${\tilde W^\pm}$ and of the charged higgsinos ${\tilde H_1^-}$,
${\tilde H_2^+}$. Their mass terms are given by
\begin{equation}
  \left( \matrix{ {\tilde W^-} & {\tilde H_1^-} } \right)
  \> {\cal M}_{\tilde{\chi}^\pm} \>
  \left( \matrix{ {\tilde W^+} \cr {\tilde H_2^+} } \right)
  + \mbox{\rm h.c.} 
\end{equation}
Their mass matrix,
\begin{eqnarray}
  {\cal M}_{\tilde{\chi}^\pm} = 
  \left( \matrix{
  {M_2} & {gv_2} \cr
  {gv_1} & \mu 
  } \right) ,
\end{eqnarray}
is diagonalized by the following linear combinations
\begin{eqnarray}
  \tilde{\chi}^-_i & = & U_{i1} \tilde{W}^- + U_{i2} \tilde{H}_1^- , 
\\
  \tilde{\chi}^+_i & = & V_{i1} \tilde{W}^+ + V_{i2} \tilde{H}_2^+ .
\end{eqnarray}
We choose ${\rm det}(U)=1$ and $U^* {\cal M}_{\tilde{\chi}^\pm}
V^\dagger = {\rm diag} ( m_{\tilde{\chi}^\pm_1},
m_{\tilde{\chi}^\pm_2} )$ with non-negative chargino masses $
m_{\tilde{\chi}^\pm_i} \ge 0$.
We do not include any one-loop corrections to the chargino masses 
since they are negligible compared to the corrections $\delta_{33}$ 
and $\delta_{44}$ introduced above for the neutralino masses 
\cite{NeuLoop1}.
 
When discussing the squark mass matrix including mixing, it is
convenient to choose a basis where the squarks are rotated in the same
way as the corresponding quarks in the standard model.  We follow the
conventions of the particle data group \cite{PDG} and put the mixing
in the left-handed $d$-quark fields, so that the definition of the
Cabibbo-Kobayashi-Maskawa matrix is $\mbox{\bf K}= \mbox{\bf V}_1
\mbox{\bf V}_2^\dagger$, where $\mbox{\bf V}_1$ ($\mbox{\bf V}_2$)
rotates the interaction left-handed $u$-quark ($d$-quark) fields to
mass eigenstates.  For sleptons we choose an analogous basis, but due
to the masslessness of neutrinos no analog of the CKM matrix appears.
 
We then obtain the general $6\times6$ $\tilde{u}$- and
$\tilde{d}$-squark mass matrices:
\begin{equation}
  {\cal M}_{\tilde u}^2 = \left( \matrix{
  \mbox{\bf M}_Q^2 + \mbox{\bf m}_u^\dagger \mbox{\bf m}_u +
      D_{LL}^{u} \mbox{\bf 1} &
   \mbox{\bf m}_u^\dagger 
        ( {\bf A}_U^\dagger - \mu^* \cot\beta ) \cr
   ( {\bf A}_U - \mu \cot\beta ) \mbox{\bf m}_u &
  \mbox{\bf M}_U^2 + \mbox{\bf m}_u \mbox{\bf m}_u^\dagger +
      D_{RR}^{u} \mbox{\bf 1} \cr
  } \right),
  \label{mutilde}
\end{equation}
\begin{equation}
  {\cal M}_{\tilde d}^2=\left( {\matrix{
  {\mbox{\bf K}^\dagger \mbox{\bf M}_Q^2 \mbox{\bf K}+
  \mbox{\bf m}_d\mbox{\bf m}_d^\dagger+D_{LL}^{d}\mbox{\bf 1}}&
  {\mbox{\bf m}_d^\dagger ( {\bf A}_D^\dagger-\mu^*\tan\beta )}\cr
  {( {\bf A}_D-\mu\tan\beta ) \mbox{\bf m}_d}&
  {\mbox{\bf M}_D^2+\mbox{\bf m}_d^\dagger\mbox{\bf m}_d+
      D_{RR}^{d}\mbox{\bf 1}}\cr
  }} \right),
  \label{mdtilde}
\end{equation}
and the general sneutrino and charged slepton masses
\begin{equation}
  {\cal M}^2_{\tilde\nu} = \mbox{\bf M}_L^2 + D^\nu_{LL} \mbox{\bf 1}
\end{equation}
\begin{equation}
  {\cal M}^2_{\tilde e} =\left( {\matrix{
  {\mbox{\bf M}_L^2+\mbox{\bf m}_e\mbox{\bf m}_e^\dagger+
       D_{LL}^{e}\mbox{\bf 1}}&
  {\mbox{\bf m}_e^\dagger ( {\bf A}_E^\dagger-\mu^*\tan\beta )}\cr
  {( {\bf A}_E-\mu\tan\beta ) \mbox{\bf m}_e}&
  {\mbox{\bf M}_E^2+\mbox{\bf m}_e^\dagger\mbox{\bf m}_e+
       D_{RR}^{e}\mbox{\bf 1}}\cr
  }} \right).
  \label{metilde}
\end{equation}
Here
\begin{equation}
  D^f_{LL}=m_Z^2\cos 2\beta(T_{3f}-e_f\sin^2\theta_W),
\end{equation}
\begin{equation}
  D^f_{RR}=m_Z^2\cos 2\beta e_f\sin^2\theta_W
\end{equation}
where $T_{3f}$ is the third component of the weak isospin and $e_{f}$ 
is the charge in units of the absolute value of the electron charge, $e$.
In the chosen basis, we have $\mbox{\bf m}_u$ = $\mbox{\rm diag} 
\left( m_{\rm
u}, m_{\rm c}, m_{\rm t} \right)$, $\mbox{\bf m}_d $ = $\mbox{\rm 
diag}
\left(m_{\rm d}, m_{\rm s}, m_{\rm b} \right)$ and $\mbox{\bf m}_e $ 
= $
\mbox{\rm diag} (m_e, m_\mu, m_\tau )$.

The slepton and squark mass eigenstates $\tilde{f}_k$ ($\tilde{\nu}_k$
with $k=1,2,3$ and $\tilde{e}_k$, $\tilde{u}_k$ and $\tilde{d}_k$ with
$k=1,\dots,6$) diagonalize the previous mass matrices and are related
to the current sfermion eigenstates $\tilde{f}_{La}$ and 
$\tilde{f}_{Ra}$ ($a=1,2,3$) via 
\begin{eqnarray}
  \tilde{f}_{La} & = & \sum_{k=1}^6 \tilde{f}_k {\bf 
\Gamma}_{FL}^{*ka} , \\
  \tilde{f}_{Ra} & = & \sum_{k=1}^6 \tilde{f}_k {\bf 
\Gamma}_{FR}^{*ka} .
\end{eqnarray} 
The squark and charged slepton mixing matrices ${\bf \Gamma}_{UL,R}$,
${\bf \Gamma}_{DL,R}$ and ${\bf \Gamma}_{EL,R}$ have dimension
$6\times 3$, while the sneutrino mixing matrix ${\bf \Gamma}_{\nu L}$
has dimension $3\times3$.

For simplicity we make a simple Ansatz for the up-to-now
arbitrary soft supersymmetry-breaking parameters:
\begin{equation}
  \begin{array}{l}
  {\bf A}_U = \mbox{\rm diag}(0,0,A_t) \\ 
  {\bf A}_D = \mbox{\rm diag}(0,0,A_b) \\
  {\bf A}_E = 0 \\
  \mbox{\bf M}_Q = \mbox{\bf M}_U = \mbox{\bf M}_D = \mbox{\bf M}_E =
   \mbox{\bf M}_L = m_0 \mbox{\bf 1} .
  \end{array}
  \label{ansatz}
\end{equation}
This allows the squark mass matrices to be diagonalized analytically.
For example, for the top squark one has, in terms of the top squark 
mixing
angle $\theta_{\tilde{t}}$,
\begin{equation}
  \Gamma_{UL}^{\tilde{t}_1\tilde{t}} =
  \Gamma_{UR}^{\tilde{t}_2\tilde{t}} = \cos \theta_{\tilde{t}} ,
  \qquad
  \Gamma_{UL}^{\tilde{t}_2\tilde{t}} =
  - \Gamma_{UR}^{\tilde{t}_1\tilde{t}} = \sin \theta_{\tilde{t}} .
\end{equation}

Notice that the Ansatz (\ref{ansatz}) implies the absence of
tree-level flavor changing neutral currents in all sectors of the
model. 


\section{The Boltzmann equation and thermal averaging}
\label{sec:Boltzmann}

Griest and Seckel \cite{GriestSeckel} have worked out the Boltzmann
equation when coannihilations are included. We start by reviewing
their expressions and then continue by rewriting them into a more
convenient form that resembles the familiar case without
coannihilations. This allows us to use similar expressions for
calculating thermal averages and solving the Boltzmann equation
whether coannihilations are included or not.

\subsection{Review of the Boltzmann equation with coannihilations}

Consider annihilation of $N$ supersymmetric particles $\chi_i$
($i=1,\ldots,N$) with masses $m_i$ and internal degrees of freedom
(statistical weights) $g_i$.  Also assume that $m_1 \leq m_2 \leq
\cdots \leq m_{N-1} \leq m_N$ and that $R$-parity is conserved. Note
that for the mass of the lightest neutralino we will use the
notation $m_{\chi}$ and $m_{1}$ interchangeably.

The evolution of the number density $n_i$ of particle $i$ is
\begin{eqnarray} \label{eq:Boltzmann}
  \frac{dn_{i}}{dt} 
  &=& 
  -3 H n_{i} 
  - \sum_{j=1}^N \langle \sigma_{ij} v_{ij} \rangle 
    \left( n_{i} n_{j} - n_{i}^{\rm{eq}} n_{j}^{\rm{eq}} \right) 
  \nonumber \\ 
  & & 
  - \sum_{j\ne i} 
  \big[ \langle \sigma'_{Xij} v_{ij} \rangle 
        \left( n_i n_X - n_{i}^{\rm{eq}} n_{X}^{\rm{eq}} \right)
      - \langle \sigma'_{Xji} v_{ij} \rangle
        \left( n_j n_X - n_{j}^{\rm{eq}} n_{X}^{\rm{eq}} \right)
  \big]
  \nonumber \\ 
  & &
  - \sum_{j\ne i} 
  \big[ \Gamma_{ij} 
        \left( n_i - n_i^{\rm{eq}} \right) 
      - \Gamma_{ji} 
        \left( n_j - n_j^{\rm{eq}} \right) 
  \big] .
\end{eqnarray}
The first term on the right-hand side is the dilution due to the
expansion of the Universe. $H$ is the Hubble parameter. The second
term describes $\chi_i\chi_j$ annihilations, whose total
annihilation cross section is 
\begin{eqnarray}
  \sigma_{ij}  & = & \sum_X \sigma (\chi_i \chi_j \rightarrow X).
\end{eqnarray}
The third term describes $\chi_i \to \chi_j$ conversions by
scattering off the cosmic thermal background,
\begin{eqnarray}
  \sigma'_{Xij} & = & \sum_Y \sigma (\chi_i X \rightarrow \chi_j Y)
\end{eqnarray}
being the inclusive scattering cross section. The last term accounts
for $\chi_i$ decays, with inclusive decay rates 
\begin{eqnarray}
  \Gamma_{ij}  & = & \sum_X \Gamma (\chi_i \rightarrow \chi_j X).
\end{eqnarray}
In the previous expressions, $X$ and $Y$
are (sets of) standard model particles involved in the
interactions, $v_{ij}$ is the `relative velocity' defined by
\begin{equation}
  v_{ij} = \frac{\sqrt{(p_{i} \cdot p_{j})^2-m_{i}^2 m_{j}^2}}{E_{i} E_{j}}
\end{equation}
with $p_{i}$ and $E_{i}$ being the four-momentum and energy of 
particle $i$, and finally $n_{i}^{\rm{eq}}$ is the equilibrium number
density of particle $\chi_i$,
\begin{equation}
  n_{i}^{\rm{eq}} = \frac{g_{i}}{(2\pi)^3} \int d^3{\bf p}_{i}f_{i}
\end{equation}
where ${\bf p}_i$ is the three-momentum of particle $i$, and
 $f_i$ is its equilibrium distribution function. 
In the Maxwell-Boltzmann approximation it is given by
\begin{equation}
  f_{i} = e^{-E_{i}/T}.
\end{equation}
The thermal average $\langle\sigma_{ij}v_{ij}\rangle$ is defined
with equilibrium distributions and is given by
\begin{equation}
  \langle \sigma_{ij}v_{ij} \rangle = \frac{\int d^3{\bf
      p}_{i}d^3{\bf p}_{j} 
  f_{i}f_{j}\sigma_{ij}v_{ij}}
  {\int d^3{\bf p}_{i}d^3{\bf p}_{j}f_{i}f_{j}}
\end{equation}

Normally, the decay rate of supersymmetric particles $\chi_i$ other
than the lightest which is stable is much faster than the age of the
universe. Since we have assumed $R$-parity conservation, all of these
particles decay into the lightest one. So its final abundance is
simply described by the sum of the density of all supersymmetric
particles,
\begin{equation}
  n= \sum_{i=1}^N n_{i}.
\end{equation}
For $n$ we get the following evolution equation
\begin{equation}
  \frac{dn}{dt} = -3Hn - \sum_{i,j=1}^N \langle \sigma_{ij} v_{ij} \rangle 
  \left( n_{i}n_{j} - n_{i}^{\rm{eq}}n_{j}^{\rm{eq}} \right)
\end{equation}
where the terms on the second and third lines in
Eq.~(\ref{eq:Boltzmann}) cancel in the sum. 

The scattering rate of supersymmetric particles off particles in the
thermal background is much faster than their annihilation rate,
because the scattering cross sections $\sigma'_{Xij}$ are of the same
order of magnitude as the annihilation cross sections $\sigma_{ij}$
but the background particle density $n_X$ is much larger than each of
the supersymmetric particle densities $n_i$ when the former are
relativistic and the latter are non-relativistic, and so suppressed by
a Boltzmann factor. In this case, the $\chi_i$ distributions remain in
thermal equilibrium, and in particular their ratios are equal to the
equilibrium values,
\begin{equation}
  \frac{n_{i}}{n} \simeq \frac{n_{i}^{\rm{eq}}}{n^{\rm{eq}}}.
\end{equation}
We then get
\begin{equation} \label{eq:Boltzmann2}
  \frac{dn}{dt} =
  -3Hn - \langle \sigma_{\rm{eff}} v \rangle 
  \left( n^2 - n_{\rm{eq}}^2 \right)
\end{equation}
where
\begin{equation} \label{eq:sigmaveffdef}
  \langle \sigma_{\rm{eff}} v \rangle = \sum_{ij} \langle
  \sigma_{ij}v_{ij} \rangle \frac{n_{i}^{\rm{eq}}}{n^{\rm{eq}}}
  \frac{n_{j}^{\rm{eq}}}{n^{\rm{eq}}}.
\end{equation}


\subsection{Thermal averaging}
\label{sec:thermav}

So far the reviewing. Now let's continue
by reformulating the thermal averages into
more convenient expressions. 

We rewrite Eq.~(\ref{eq:sigmaveffdef}) as
\begin{equation} \label{eq:sigmaveff}
  \langle \sigma_{\rm{eff}} v \rangle = \frac{ \sum_{ij} \langle
  \sigma_{ij}v_{ij} \rangle n_{i}^{\rm{eq}} n_{j}^{\rm{eq}}}{n^2_{\rm{eq}}}
  = 
  \frac{A}{n_{\rm{eq}}^2} \, .
\end{equation}

For the denominator we obtain, 
using Boltzmann statistics for $f_i$,
\begin{equation} \label{eq:neq}
  n^{\rm eq} = \sum_i n^{\rm eq}_i = 
  \sum_i \frac{g_i}{(2\pi)^3} \int d^3 p_i 
  e^{-E_{i}/T} = 
  \frac{T}{2\pi^2} \sum_i g_i m_{i}^2
  K_{2} \left( \frac{m_{i}}{T}\right)
\end{equation}
where $K_{2}$ is the modified Bessel function of the second kind of 
order 2.

The numerator is the total annihilation rate per unit volume
at temperature $T$,
\begin{equation} 
  A = \sum_{ij} \langle \sigma_{ij} v_{ij} \rangle n_i^{\rm eq}
  n_j^{\rm eq} = \sum_{ij} \frac{g_{i}g_{j}}{(2\pi)^6} \int d^3 {\bf p}_{i}
  d^3{\bf p}_{j} f_{i}f_{j} \sigma_{ij} v_{ij}
\end{equation}
It is convenient
to cast it in a covariant form,
\begin{equation} 
  A = \sum_{ij} 
  \int W_{ij} \frac{g_i f_i d^3{\bf p}_i}{(2\pi)^3 2E_i}
  \frac{g_j f_j d^3{\bf p}_j}{(2\pi)^3 2E_j} .
\label{eq:Aij2}
\end{equation}
$W_{ij}$ is the (unpolarized) annihilation rate per unit volume
corresponding to the covariant normalization of $2E$ colliding
particles per unit volume. $W_{ij}$ is a dimensionless Lorentz
invariant, related to the (unpolarized) cross section
through\footnote{The quantity $w_{ij}$ in Ref.\ \protect\cite{SWO}
  is $W_{ij}/4$.}
\begin{equation} \label{eq:Wijcross}
  W_{ij} = 4 p_{ij} \sqrt{s} \sigma_{ij} = 4 \sigma_{ij} \sqrt{(p_i
\cdot p_j)^2 - m_i^2 m_j^2} = 4 E_{i} E_{j} \sigma_{ij} v_{ij} .
\end{equation}
Here
\begin{equation}
  p_{ij} =
\frac{\left[s-(m_i+m_j)^2\right]^{1/2}
\left[s-(m_i-m_j)^2\right]^{1/2}}{2\sqrt{s}}
\end{equation}
is the momentum of particle $\chi_i$ (or $\chi_j$) in the
center-of-mass frame of the pair $\chi_i\chi_j$.

Averaging over initial and summing over final internal states, the
contribution to $W_{ij}$ of a general $n$-body final state is
\begin{equation}
  W^{n\rm{-body}}_{ij} = 
  \frac{1}{g_i g_j S_f} \sum_{\rm{internal~d.o.f.}} 
  \int  \left| {\cal M} \right|^2 (2\pi)^4 
\delta^4(p_i+p_j-{\textstyle \sum_f}p_f) \prod_f
   \frac{d^3{\bf p}_f}{(2\pi)^3 2E_f} , 
\end{equation}
where $S_f$ is a symmetry factor accounting for identical final state
particles (if there are $K$ sets of $N_k$ identical particles,
$k=1,\dots,K$, then $S_f = \prod_{k=1}^{K} N_k!$).  In particular, 
the contribution
of a two-body final state can be written as
\begin{equation}
  W^{\rm{2-body}}_{ij\to kl} = \frac{p_{kl}}{16\pi^2 g_i g_j S_{kl} \sqrt{s}}
  \sum_{\rm{internal~d.o.f.}} \int \left| {\cal M}(ij\to kl) \right|^2
  d\Omega ,
\end{equation}
where $p_{kl}$ is the final center-of-mass momentum, $S_{kl}$ is a
symmetry factor equal to 2 for identical final particles and to 1
otherwise, and the integration is over the outgoing directions of
one of the final particles.  As usual, an average over initial
internal degrees of freedom is performed.

We now reduce the integral in the covariant expression for $A$,
Eq.~(\ref{eq:Aij2}), from 6 dimensions to 1.
Using Boltzmann statistics for $f_i$ (a good approximation for
$T\lsim m$)
\begin{equation} \label{eq:Aij2b}
  A =
  \sum_{ij} \int g_i g_j W_{ij} e^{-E_{i}/T} e^{-E_{j}/T} 
\frac{d^3{\bf p}_i}{(2\pi)^3 2E_i}
  \frac{d^3{\bf p}_j}{(2\pi)^3 2E_j} ,
\end{equation}
where ${\bf p}_{i}$ and ${\bf p}_{j}$ are the three-momenta and
$E_{i}$ and $E_{j}$ are the energies of the colliding particles.
Following the procedure in Ref.~\cite{GondoloGelmini} we then rewrite
the momentum volume element as
\begin{equation}
  d^3 {\bf p}_{i} d^3 {\bf p}_{j} = 4 \pi |{\bf p}_{i}| E_i dE_{i}
  \, 4 \pi |{\bf p}_{j}| E_j dE_{j} \, \frac{1}{2} d \cos \theta
\end{equation}
where $\theta$ is the angle between ${\bf p}_{i}$ and 
${\bf p}_{j}$. Then we change integration variables from 
$E_{i}$, $E_{j}$, $\theta$ to $E_{+}$, $E_{-}$ and $s$, given by
\begin{equation}
  \left\{ \begin{array}{lcl}
  E_{+} & = & E_{i}+E_{j} \\
  E_{-} & = & E_{i}-E_{j} \\
  s & = & m_{i}^2+m_{j}^2 + 2E_{i}E_{j}-2 |{\bf p}_{i}| |{\bf
    p}_{j}| \cos \theta,
  \end{array} \right.
\end{equation}
whence the volume element becomes
\begin{equation}
  \frac{d^3{\bf p}_i}{(2\pi)^3 2E_i} \frac{d^3{\bf p}_j}{(2\pi)^3 2E_j} =
  \frac{1}{(2\pi)^4} \frac{dE_{+}dE_{-}ds}{8},
\end{equation}
and the integration region $\{ E_i \geq m_i, E_j \geq m_j, |\cos \theta| 
\leq
1\}$ transforms into 
\begin{eqnarray}
  && s \geq (m_i+m_j)^2, \\ && E_{+} \geq \sqrt{s} , \\ && \left\vert
  E_{-} - E_{+} \frac{m_j^2-m_i^2}{s} \right\vert \leq 2 p_{ij}
  \sqrt{\frac{E_{+}^2-s}{s}}.
\end{eqnarray}

Notice now that the product of the equilibrium distribution
functions depends only on $E_{+}$ and not $E_{-}$ due to the
Maxwell-Boltzmann approximation, and that the invariant rate
$W_{ij}$ depends only on $s$ due to the neglect of final state
statistical factors. Hence we can immediately integrate over
$E_{-}$,
\begin{equation}
  \int dE_{-} = 4p_{ij} \sqrt{\frac{E_{+}^2-s}{s}}.
\end{equation}
The volume element is now
\begin{equation}
  \frac{d^3{\bf p}_i}{(2\pi)^3 2E_i} \frac{d^3{\bf p}_j}{(2\pi)^3 2E_j} = 
  \frac{1}{(2\pi)^4} \frac{p_{ij}}{2} \sqrt{\frac{E_{+}^2-s}{s}} 
dE_{+} ds 
\end{equation}

We now perform the $E_{+}$ integration. We obtain
\begin{equation}
\label{eq:As}
  A = \frac{T}{32 \pi^4} \sum_{ij} \int_{(m_i+m_j)^2}^\infty ds
  g_ig_jp_{ij} W_{ij} K_{1} \left( \frac{\sqrt{s}}{T}\right)
\end{equation}
where $K_{1}$ is the modified Bessel function of the second kind of 
order 1.

We can take the sum inside the integral and define an effective
annihilation rate $W_{\rm eff}$ through
\begin{equation}
  \sum_{ij} g_i g_j p_{ij} W_{ij} = g_1^2 p_{\rm{eff}} W_{\rm{eff}}
\end{equation}
with
\begin{equation}
\label{eq:peff}
  p_{\rm{eff}} = p_{11} = \frac{1}{2} \sqrt{s-4m_{1}^2} .
\end{equation}
In other words
\begin{equation} \label{eq:weff}
  W_{\rm{eff}} = \sum_{ij}\frac{p_{ij}}{p_{11}}
  \frac{g_ig_j}{g_1^2} W_{ij} = 
  \sum_{ij} \sqrt{\frac{[s-(m_{i}-m_{j})^2][s-(m_{i}+m_{j})^2]}
  {s(s-4m_1^2)}} \frac{g_ig_j}{g_1^2} W_{ij}.
\end{equation}
Because $W_{ij}(s) = 0 $ for $s \le (m_i+m_j)^2$, the radicand is  
never negative.

In terms of cross sections, this is equivalent to the definition
\begin{equation}
\sigma_{\rm eff} = \sum_{ij} \frac{p^2_{ij}}{p^2_{11}}
  \frac{g_ig_j}{g_1^2} \sigma_{ij}.
\end{equation}  

Eq.~(\ref{eq:As}) then reads
\begin{equation}
  A = \frac{g_1^2 T}{32 \pi^4} \int_{4m_1^2}^\infty ds
  p_{\rm eff} W_{\rm eff} K_{1} \left( \frac{\sqrt{s}}{T}\right)
\end{equation}
This can be written in a form more suitable
for numerical integration by using $p_{\rm{eff}}$ instead of $s$ as
integration variable.  From Eq.~(\ref{eq:peff}), we have 
 $ ds = 8 p_{\rm{eff}} dp_{\rm{eff}} $, and 
\begin{equation}
\label{eq:Apeff}
  A = \frac{g_1^2 T}{4 \pi^4} \int_{0}^\infty dp_{\rm eff}
  p^2_{\rm eff} W_{\rm eff} K_{1} 
  \left( \frac{\sqrt{s}}{T}\right)
\end{equation}
with
\begin{equation}
  s = 4p_{\rm{eff}}^2 + 4m_1^2
\end{equation}
So we have succeeded in rewriting $A$ as a 1-dimensional integral.

{}From Eqs.~(\ref{eq:Apeff}) and~(\ref{eq:neq}), the thermal average of
the effective cross section results
\begin{equation} \label{eq:sigmavefffin2}
  \langle \sigma_{\rm{eff}}v \rangle = \frac{\int_0^\infty
  dp_{\rm{eff}} p_{\rm{eff}}^2 W_{\rm{eff}} K_1 \left(
  \frac{\sqrt{s}}{T} \right) } { m_1^4 T \left[ \sum_i \frac{g_i}{g_1}
  \frac{m_i^2}{m_1^2} K_2 \left(\frac{m_i}{T}\right) \right]^2}.
\end{equation}
This expression is very similar to the case without coannihilations,
the differences being the denominator and the replacement of the
annihilation rate with the effective annihilation rate. 
In the absence of coannihilations, this expression
correctly reduces to the formula in Gondolo and
Gelmini~\cite{GondoloGelmini}.

The definition of an effective annihilation rate independent of
temperature is a remarkable calculational advantage. As in the case
without coannihilations, the effective annihilation rate can in fact
be tabulated in advance, before taking the thermal average and
solving the Boltzmann equation.

\begin{figure}
  \centerline{\epsfig{file=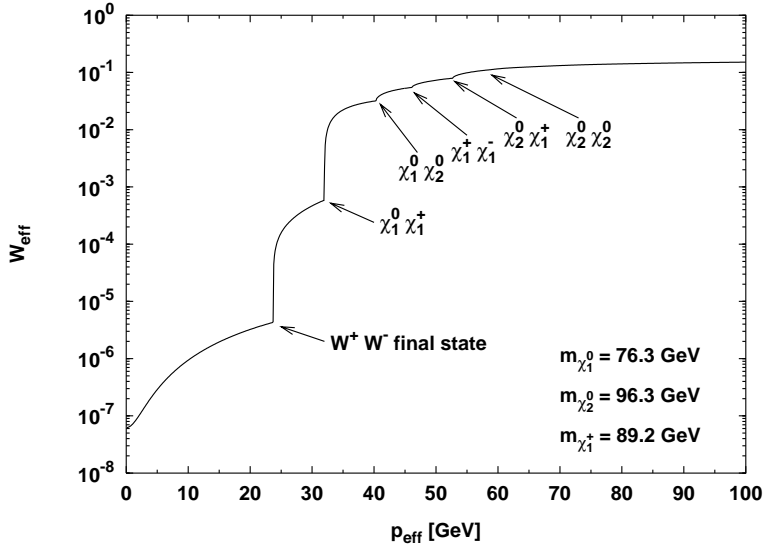,width=0.75\textwidth}}
  \caption{The effective invariant annihiliation rate $W_{\rm eff}$
    as a function of $p_{\rm eff}$ for model 1 in
    Table~\protect\ref{tab:reprmod}. The final state threshold for
    annihilation into $W^+ W^-$ and the coannihilation thresholds, as
    given by Eq.~(\protect\ref{eq:weff}), are indicated.  
    The $\chi_2^0 \chi_2^0$ coannihilation threshold is too small to
    be seen.}
  \label{fig:effrate}
\end{figure}

\begin{figure}
  \centerline{\epsfig{file=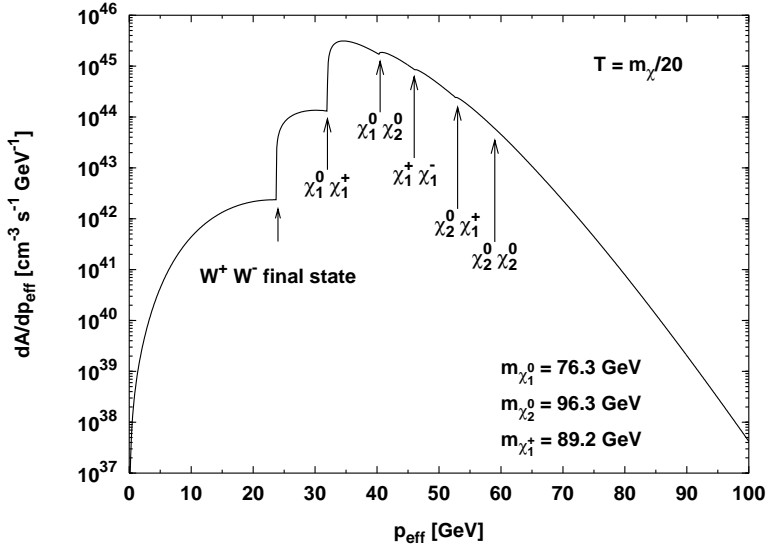,width=0.75\textwidth}}
  \caption{Total differential annihilation rate per unit volume 
    $dA/dp_{\rm eff}$ for the same model as in
    Fig.~\protect\ref{fig:effrate}, evaluated at a temperature
    $T=m_\chi/20$, typical of freeze-out. Notice the Boltzmann
    suppression at high $p_{\rm eff}$.}
  \label{fig:k1effrate}
\end{figure}

In the effective annihilation rate, coannihilations appear
as thresholds at $\sqrt{s}$ equal to the sum of the masses of the
coannihilating particles.  We show an example in
Fig.~\ref{fig:effrate} where it is clearly seen that the
coannihilation thresholds appear in the effective invariant rate
just as final state thresholds do.  For the same example,
Fig.~\ref{fig:k1effrate} shows the differential annihilation rate
per unit volume $dA/dp_{\rm eff}$, the integrand in
Eq.~(\ref{eq:Apeff}), as a function of $p_{\rm eff}$. We have
chosen a temperature $T=m_{\chi}/20$, a typical freeze-out
temperature. The Boltzmann suppression contained in the exponential
decay of $K_{1}$ at high $p_{\rm eff}$ is clearly visible.  At
higher temperatures the peak shifts to the right and at lower
temperatures to the left.  For the particular model shown in
Figs.~\ref{fig:effrate}--\ref{fig:k1effrate}, the relic density
results $\Omega_\chi h^2=0.030$ when coannihilations are included
and $\Omega_\chi h^2=0.18$ when they are not. Coannihilations
have lowered $\Omega_\chi h^2$ by a factor of 6.

We end this section with a comment on the internal degrees of
freedom $g_i$.  A neutralino is a Majorana fermion and has two
internal degrees of freedom, $g_{\chi_{i}^0}=2$. A chargino can be
treated either as two separate species $\chi_{i}^+$ and
$\chi_{i}^-$, each with internal degrees of freedom
$g_{\chi^+}=g_{\chi^-}=2$, or, more simply, as a single species
$\chi_{i}^\pm$ with $g_{\chi_{i}^\pm}=4$ internal degrees of
freedom.  The effective annihilation rates involving
charginos read
\begin{equation}
  \left\{ \begin{array}{lcl}
  W_{\chi_{i}^0 \chi_{j}^\pm} & = & W_{\chi_{i}^0 \chi_{j}^+} = 
    W_{\chi_{i}^0 \chi_{j}^-} \quad , \quad \forall\ i=1,\ldots,4,\ 
    j=1,2 \anl
  W_{\chi_{i}^\pm \chi_{j}^\pm} & = & \frac{1}{2} 
  \left[ W_{\chi_{i}^+ \chi_{j}^+} +  
  W_{\chi_{i}^+ \chi_{j}^-}\right] = 
  \frac{1}{2} \left[ W_{\chi_{i}^- \chi_{j}^-} +  
  W_{\chi_{i}^- \chi_{j}^+}\right] \quad , \quad \forall\ i,j=1,2
  \end{array} \right.
\end{equation}

\subsection{Reformulation of the Boltzmann equation}

We now follow Gondolo and Gelmini \cite{GondoloGelmini} to 
put Eq.~(\ref{eq:Boltzmann2}) in a more convenient form by
considering the ratio of the number density to the entropy density,
\begin{equation} \label{eq:ydef}
  Y = \frac{n}{s}.
\end{equation}
Consider
\begin{equation}
  \frac{dY}{dt} = \frac{d}{dt} \left( \frac{n}{s} \right) = 
  \frac{\dot{n}}{s}-\frac{n}{s^2}\dot{s}
\end{equation}
where dot means time derivative. In absence
of entropy production, $S=R^3s$ is constant ($R$ is the scale factor).
Differentiating with respect to time we see 
that
\begin{equation}
  \dot{s} = -3\frac{\dot{R}}{R} s = -3Hs
\label{eq:entropycons}
\end{equation}
which yields
\begin{equation}
  \dot{Y} = \frac{\dot{n}}{s} + 3H \frac{n}{s}.
\end{equation}
Hence we can rewrite Eq.~(\ref{eq:Boltzmann2}) as
\begin{equation} \label{eq:Boltzmann3}
  \dot{Y} = -s  \langle \sigma_{\rm{eff}} v \rangle 
  \left( Y^2 - Y_{\rm{eq}}^2 \right).
\end{equation}

The right-hand side depends only on temperature, and it is therefore
convenient to use temperature $T$ instead of time $t$ as independent
variable. Defining $x=m_1/T$ we have
\begin{equation}
  \frac{dY}{dx} = - \frac{m_{1}}{x^2} \frac{1}{3H} \frac{ds}{dT}
  \langle \sigma_{\rm{eff}} v \rangle \left( Y^2 -
  Y_{\rm{eq}}^2 \right).
\label{eq:Boltzmann3bis}
\end{equation}
where we have used
\begin{equation}
  \frac{1}{\dot{T}} = \frac{1}{\dot{s}} \frac{ds}{dT} = -
  \frac{1}{3Hs} \frac{ds}{dT} 
\end{equation} 
which follows from Eq.~(\ref{eq:entropycons}). 
With the Friedmann equation in a radiation dominated universe
\begin{equation}
  H^2 = \frac{8\pi G \rho}{3} ,
\end{equation}
where $G$ is the gravitational constant, and the
usual parameterization of the energy and entropy densities
in terms of the effective degrees of freedom $g_{\rm{eff}}$ and
$h_{\rm{eff}}$, \begin{equation} \label{eq:geffheff}
  \rho = g_{\rm{eff}}(T) \frac{\pi^2}{30} T^4
  , \quad 
  s = h_{\rm{eff}}(T) \frac{2\pi^2}{45} T^3 ,
\end{equation}
we can cast Eq.~(\ref{eq:Boltzmann3bis})
into the form \cite{GondoloGelmini}
\begin{equation} \label{eq:Boltzmann4}
  \frac{dY}{dx} = - \sqrt{\frac{\pi}{45G}} \frac{g_{*}^{1/2}m_1}{x^2}
  \langle \sigma_{\rm{eff}} v \rangle \left( Y^2 -
  Y_{\rm{eq}}^2 \right) 
\end{equation}
where $Y_{\rm eq}$ can be written as
\begin{equation}
  Y_{\rm{eq}} = \frac{n_{\rm{eq}}}{s} = 
  \frac{45 x^2}{4 \pi^4 h_{\rm{eff}}(T)} \sum_i g_i
  \left( \frac{m_i}{m_1} \right)^2 K_{2} \left( x 
\frac{m_{i}}{m_1}\right),
\end{equation}
using Eqs.~(\ref{eq:neq}), (\ref{eq:ydef}) and
(\ref{eq:geffheff}).

The parameter $g_{*}^{1/2}$ is defined as
\begin{equation}
  g_{*}^{1/2} = \frac{h_{\rm{eff}}}{\sqrt{g_{\rm{eff}}}}
  \left( 1+\frac{T}{3h_{\rm{eff}}} \frac{d h_{\rm{eff}}}{dT}
  \right)
\end{equation}

For $g_{\rm eff}$, $h_{\rm eff}$ and $g_*^{1/2}$ we use the values
in Ref.~\cite{GondoloGelmini} with a QCD phase-transition
temperature $T_{QCD} = 150 $ MeV. Our results are insensitive to the
value of $T_{QCD}$, because due to a lower limit on the neutralino
mass the neutralino freeze-out temperature is always much larger
than $T_{QCD}$.

To obtain the relic density we integrate Eq.~(\ref{eq:Boltzmann4})
from $x=0$ to $x_0=m_\chi/T_0$ where $T_0$ is the photon temperature
of the Universe today. The relic density today in units of the
critical density is then given by
\begin{equation}
  \Omega_\chi = \rho_\chi^0/\rho_{\rm
  crit}=m_\chi s_0 Y_0/\rho_{\rm crit}
\end{equation}
where $\rho_{\rm crit}=3 H^2/8 \pi G$ is the critical density, $s_{0}$ 
is the entropy density today and $Y_{0}$ is the result of the 
integration of Eq.~(\ref{eq:Boltzmann4}). With a
background radiation temperature of $T_0=2.726$ K we finally obtain
\begin{equation} \label{eq:omegah2}
  \Omega_\chi h^2 = 2.755\times 10^8 \frac{m_\chi}{\mbox{GeV}} Y_0.
\end{equation}


\section{Annihilation cross sections}
\label{sec:AnnCross}

\begin{table}
\begin{center}
\begin{tabular}{lll} \hline 
  Initial state & Final state & Feynman diagrams \\ \hline
   & $H_1 H_1$, $H_1 H_2$, $H_2 H_2$, $H_3 H_3$ &
  $t(\chi_k^0)$, $u(\chi_k^0)$, $s(H_{1,2})$ \\
   & $H_1 H_3$, $H_2 H_3$ &
  $t(\chi_k^0)$, $u(\chi_k^0)$, $s(H_{3})$, $s(Z^0)$ \\
   & $H^- H^+$ &
  $t(\chi_e^+)$, $u(\chi_e^+)$, $s(H_{1,2})$, $s(Z^0)$ \\
   & $Z^0 H_1$, $Z^0 H_2$ &
  $t(\chi_k^0)$, $u(\chi_k^0)$, $s(H_{3})$, $s(Z^0)$ \\
  $\chi_i^0 \chi_j^0$ & $Z^0 H_3$ &
  $t(\chi_k^0)$, $u(\chi_k^0)$, $s(H_{1,2})$ \\
   & $W^- H^+$, $W^+ H^-$ &
  $t(\chi_e^+)$, $u(\chi_e^+)$, $s(H_{1,2,3})$ \\
   & $Z^0 Z^0$ &
  $t(\chi_k^0)$, $u(\chi_k^0)$, $s(H_{1,2})$ \\
   & $W^- W^+$ &
  $t(\chi_e^+)$, $u(\chi_e^+)$, $s(H_{1,2})$, $s(Z^0)$ \\
   & $f \bar{f}$ &
  $t(\tilde{f}_{L,R})$, $u(\tilde{f}_{L,R})$, $s(H_{1,2,3})$,
  $s(Z^0)$ \\ \hline  
   & $H^+ H_1$, $H^+ H_2$ &
  $t(\chi_k^0)$, $u(\chi_e^+)$, $s(H^+)$, $s(W^+)$ \\
   & $H^+ H_3$ &
  $t(\chi_k^0)$, $u(\chi_e^+)$, $s(W^+)$ \\
   & $W^+ H_1$, $W^+ H_2$ &
  $t(\chi_k^0)$, $u(\chi_e^+)$, $s(H^+)$, $s(W^+)$ \\
   & $W^+ H_3$ &
  $t(\chi_k^0)$, $u(\chi_e^+)$, $s(H^+)$ \\
  $\chi_c^+ \chi_i^0$ & $H^+ Z^0$ &
  $t(\chi_k^0)$, $u(\chi_e^+)$, $s(H^+)$ \\
   & $\gamma H^+$ &
  $t(\chi_c^+)$, $s(H^+)$ \\
   & $W^+ Z^0$ &
  $t(\chi_k^0)$, $u(\chi_e^+)$, $s(W^+)$ \\
   & $\gamma W^+$ &
  $t(\chi_c^+)$, $s(W^+)$ \\
   & $u \bar{d}$ &
  $t(\tilde{d}_{L,R})$, $u(\tilde{u}_{L,R})$, $s(H^+)$, $s(W^+)$ \\
   & $\nu \bar{\ell}$ &
  $t(\tilde{\ell}_{L,R})$, $u(\tilde{\nu}_{L})$, $s(H^+)$, $s(W^+)$
  \\ \hline
   & $H_1 H_1$, $H_1 H_2$, $H_2 H_2$, $H_3 H_3$ &
  $t(\chi_e^+)$, $u(\chi_e^+)$, $s(H_{1,2})$ \\
   & $H_1 H_3$, $H_2 H_3$ &
  $t(\chi_e^+)$, $u(\chi_e^+)$, $s(H_{3})$, $s(Z^0)$ \\
   & $H^+ H^-$ &
  $t(\chi_k^0)$, $s(H_{1,2})$, $s(Z^0,\gamma)$ \\
   & $Z^0 H_1$, $Z^0 H_2$ &
  $t(\chi_e^+)$, $u(\chi_e^+)$, $s(H_{3})$, $s(Z^0)$ \\
   & $Z^0 H_3$ &
  $t(\chi_e^+)$, $u(\chi_e^+)$, $s(H_{1,2})$ \\
   & $H^+ W^-$, $W^+ H^-$ &
  $t(\chi_e^+)$, $s(H_{1,2,3})$ \\
  $\chi_c^+ \chi_d^-$ & $Z^0 Z^0$ &
  $t(\chi_e^+)$, $u(\chi_e^+)$, $s(H_{1,2})$ \\
   & $W^+ W^-$ &
  $t(\chi_k^0)$, $s(H_{1,2})$, $s(Z^0, \gamma)$ \\
   & $\gamma \gamma$ (only for $c=d$) &
  $t(\chi_c^+)$, $u(\chi_c^+)$ \\
   & $Z^0 \gamma$ &
  $t(\chi_d^+)$, $u(\chi_c^+)$ \\
   & $u \bar{u}$ &
  $t(\tilde{d}_{L,R})$, $s(H_{1,2,3})$, $s(Z^0, \gamma)$ \\
   & $\nu \bar{\nu}$ &
  $t(\tilde{\ell}_{L,R})$, $s(Z^0)$ \\
   & $\bar{d} d$ &
  $t(\tilde{u}_{L,R})$, $s(H_{1,2,3})$, $s(Z^0, \gamma)$ \\
   & $\bar{\ell} \ell$ &
  $t(\tilde{\nu}_{L})$, $s(H_{1,2,3})$, $s(Z^0, \gamma)$ \\ \hline
   & $H^+ H^+$ &
  $t(\chi_k^0)$, $u(\chi_k^0)$ \\
  $\chi_c^+ \chi_d^+$ & $H^+ W^+$ &
  $t(\chi_k^0)$, $u(\chi_k^0)$ \\
   & $W^+ W^+$ &
  $t(\chi_k^0)$, $u(\chi_k^0)$ \\ \hline
\end{tabular}
\end{center}
\caption{All Feynman diagrams for which we calculate the 
  annihilation cross section. $s(x)$, $t(x)$ and $u(x)$ denote a
  tree-level Feynman diagram in which particle $x$ is exchanged in
  the $s$-, $t$- and $u$-channel respectively. Indices $i,j,k$ run
  from 1 to 4, and indices $c,d,e$ from 1 to 2.  $u$, $\tilde{u}$,
  $d$, $\tilde{d}$, $\nu$, $\tilde{\nu}$, $\ell$, $\tilde{\ell}$,
  $f$ and $\tilde{f}$ are generic notations for up-type quarks,
  up-type squarks, down-type quarks, down-type squarks, neutrinos,
  sneutrinos, leptons, sleptons, fermions and sfermions.  A sum of
  diagrams over (s)fermion generation indices and over the
  neutralino and chargino indices $k$ and $e$ is understood (no sum
  over indices $i,j,c,d$).}
\label{tab:annchannels}
\end{table}

We have calculated all two-body final state cross sections at tree
level for neutralino-neutral\-ino, neutralino-chargino and
chargino-chargino annihilation. A complete list is given in
Table~\ref{tab:annchannels}. 

Since we have so many different diagrams contributing, we have to use 
some method where the diagrams can be calculated efficiently.  To 
achieve this, we classify diagrams according to their topology ($s$-, 
$t$- or $u$-channel) and to the spin of the particles involved.  We 
then compute the helicity amplitudes for each type of diagram 
analytically with {\sc Reduce}~\cite{reduce} using general expressions 
for the vertex couplings.  Further details will be found in 
Ref.~\cite{paolo}.

The strength of the helicity amplitude method is that the analytical
calculation of a given type of diagram has to be performed only once
and the sum of the contributing diagrams for each set of initial and
final states can be done numerically afterwards.


\section{Numerical methods}
\label{sec:NumMethods}

In this section we describe the numerical methods we use to evaluate
the effective invariant rate and its thermal average, and to integrate
the density evolution equation.

We obtain the effective invariant rate numerically as follows.  We
generate {\sc Fortran} routines for the helicity amplitudes of all
types of diagrams automatically with {\sc Reduce}, as explained in the
previous section. We sum the Feynman diagrams numerically for each
annihilation channel $ij\to kl$. We then sum the squares of the
helicity amplitudes so obtained, and sum the contributions of all
annihilation channels. Explicitly, we compute
\begin{equation} \label{eq:helsum}
  {d W_{\rm eff} \over d \cos\theta } = 
\sum_{ijkl}
{p_{ij} p_{kl} \over 32 \pi p_{\rm eff} S_{kl} \sqrt{s} }
\sum_{\rm helicities}
   \left| \sum_{\rm diagrams}  {\cal M}(ij \to kl) \right|^2
\end{equation}
where $\theta$ is the angle between particles $k$ and $i$.  (We set
$g_1=2$ as appropriate for a neutralino.)  

We integrate over $\cos\theta$ numerically by means of adaptive
gaussian integration.  In rare cases, we find resonances in the $t$-
or $u$-channels. For the process $ij\to kl$, this can occur when
$m_i < m_k$ and $m_j > m_l$ or $m_i < m_l$ and $m_j > m_k$: at
certain values of $\cos\theta$, the momentum transfer is time-like
and matches the mass of the exchanged particle.  We have regulated
the divergence by assigning a small width of a few GeV to the
neutralinos and charginos.  Our results are not sensitive to the
choice of this width.

The calculation of the effective invariant rate $W_{\rm eff}$ is the
most time-consuming part. Fortunately, thanks to the remarkable
feature of Eq.~(\ref{eq:sigmavefffin2}), $W_{\rm eff}(p_{\rm eff})$
does not depend on the temperature $T$, and it can be tabulated once
for each model.  We have to make sure that the maximum $p_{\rm eff}$
in the table is large enough to include all important resonances,
thresholds and coannihilation thresholds. In the thermal average, the
effective invariant rate is weighted by $K_1 p_{\rm eff}^2$ (see
Eq.~(\ref{eq:sigmavefffin2})). The fast exponential decay of $K_1$ at
high $p_{\rm eff}$ Boltzmann suppresses resonances and thresholds, as
we have already seen in the example in Fig.~\ref{fig:k1effrate}.  With
a typical freeze-out temperature $T=m_\chi/20$, contributions to the
thermal average from values of $p_{\rm eff}$ beyond $\sim 1.5
m_{\chi}$ are negligible, even in the most extreme case we met in
which the effective invariant rate at high $p_{\rm eff}$ was $10^{10}$
times higher than at $p_{\rm eff}=0$.  For coannihilations, this value
of $p_{\rm eff}$ corresponds to a mass of the coannihilating particle
of $\sim 1.8m_{\chi}$.  To be on the safe side all over parameter
space, we include coannihilations whenever the mass of the
coannihilating particle is less than $2.1m_\chi$, even if typically
coannihilations are important only for masses less than $1.4 m_\chi$.
For extra safety, we tabulate $W_{\rm eff}$ from $p_{\rm eff}=0$ up to
$p_{\rm eff}=20 m_\chi$, more densely in the important low $p_{\rm
  eff}$ region than elsewhere.  We further add several points around
resonances and thresholds, both explicitly and in an adaptive manner.

To perform the thermal average in Eq.~(\ref{eq:sigmavefffin2}), we
integrate over $p_{\rm eff}$ by means of adaptive gaussian
integration, using a spline to interpolate in the $(p_{\rm eff},W_{\rm
  eff})$ table. To avoid numerical problems in the integration routine
or in the spline routine, we split the integration interval at each
sharp threshold. We also explicitly check for each MSSM model that the
spline routine behaves well at thresholds and resonances.

We finally integrate the density evolution
equation~(\ref{eq:Boltzmann4}) numerically from $x=2$, where the
density still tracks the equilibrium density, to $x_0=m_\chi/T_0$. We
use an implicit trapezoidal method with adaptive stepsize. The method
is implicit because of the stiffness of the evolution equation.  The
relic density at present is then evaluated with
Eq.~(\ref{eq:omegah2}).

A more detailed description of the numerical methods will be found in
a future publication \cite{gondoloedsjo}.


\section{Selection of models}
\label{sec:SelMod}

\begin{table}
\small
\begin{center}
\begin{tabular}{lrrrrrrr} \hline
Scan & \multicolumn{1}{l}{normal} & 
\multicolumn{1}{l}{generous} & 
\multicolumn{1}{l}{light} & 
\multicolumn{1}{l}{high} & 
\multicolumn{1}{l}{high} & 
\multicolumn{1}{l}{light} &
\multicolumn{1}{l}{heavy} \\
     &          &     &
\multicolumn{1}{l}{Higgs} &
\multicolumn{1}{l}{mass 1} & 
\multicolumn{1}{l}{mass 2} &
\multicolumn{1}{l}{higgsinos} & 
\multicolumn{1}{l}{gauginos} \\ \hline
$\mu^{\rm min}$ [GeV]  & $-5000$ &$-10000$ & $-5000$ &    1000 &  $-30000$ & $-100$ & 1000 \\
$\mu^{\rm max}$ [GeV]  &  5000   &   10000 &    5000 &   30000 &    $-1000$ & 100 & 30000 \\
$M_2^{\rm min}$ [GeV]  & $-5000$ &$-10000$ & $-5000$ &    1000 &    1000 & $-1000$ & 1.9$\mu$/$-1.9\mu$ \\
$M_2^{\rm max}$ [GeV]  &    5000 &   10000 &    5000 &   30000 &   30000 & 1000 & 2.1$\mu$/$-2.1\mu$ \\
$\tan \beta^{\rm min}$ &     1.2 &     1.2 &     1.2 &     1.2 &     1.2 & 1.2 & 1.2 \\
$\tan \beta^{\rm max}$ &      50 &      50 &      50 &      50 &      50 & 2.1 & 50 \\
$m_A^{\rm min}$ [GeV]  &       0 &       0 &       0 &       0 &       0 & 0 & 0\\
$m_A^{\rm max}$ [GeV]  &    1000 &    3000 &     150 &   10000 &   10000 & 1000 & 10000 \\
$m_0^{\rm min}$ [GeV]  &     100 &     100 &     100 &    1000 &    1000 & 100 & 1000 \\
$m_0^{\rm max}$ [GeV]  &    3000 &    5000 &    3000 &   30000 &   30000 & 3000 & 30000 \\
$A_b^{\rm min}$        & $-3m_0$ & $-3m_0$ & $-3m_0$ & $-3m_0$ & $-3m_0$ & $-3m_0$ & $-3m_0$ \\
$A_b^{\rm max}$        &  $3m_0$ &  $3m_0$ &  $3m_0$ &  $3m_0$ &  $3m_0$  & $3m_0$ & $3m_0$ \\
$A_t^{\rm min}$        & $-3m_0$ & $-3m_0$ & $-3m_0$ & $-3m_0$ & $-3m_0$ & $-3m_0$ & $-3m_0$ \\
$A_t^{\rm max}$        &  $3m_0$ &  $3m_0$ &  $3m_0$ &  $3m_0$ &  $3m_0$ & $3m_0$ & $3m_0$ \\
No. of models          &    4655 &    3938 &    3342 &    1000 &    999 & 177 & 250  \\ \hline
\end{tabular}
\end{center}
\caption{The ranges of parameter values in our scans of
  supersymmetric models. For $\mu$ and $M_{2}$ the scans are uniform
  in the logarithms of the parameters and for the rest they are
  uniform in the parameters themselves. The number of models refers
  to the number of generated models satisfying experimental
  constraints.}
\label{tab:scans}
\end{table}

In Section~\ref{sec:MSSMdef} we made some simplifying assumptions to
reduce the number of parameters in the MSSM to the seven parameters
$\mu$, $M_2$, $\tan \beta$, $m_A$, $m_0$, $A_b$ and $A_t$. It is
however a non-trivial task to scan even this reduced parameter space
to a high degree of completeness. With the goal to explore a
significant fraction of parameter space, we perform many different
scans, some general and some specialized to interesting parts of
parameter space.
The ranges of
parameter values in our scans are given in
Table~\ref{tab:scans}. 

We perform a `normal' scan where we let the seven free parameters
above vary at random within wide ranges, a `generous' scan with even
more generous bounds on the parameters, a `light Higgs' scan where
we restrict to low pseudoscalar Higgs masses, and two `high mass'
scans where we explore heavy neutralinos. In addition we perform
two other special scans: one to finely sample the cosmologically
interesting light higgsino region,
the other to study heavy mixed and gaugino-like neutralinos for
which we found that coannihilations are important. 

Remember, though, that the look of our figures might change if
different scans were used. One should especially pay no attention
to the density of points in different regions: it is just an artifact
of our scanning.

We keep only models that satisfy the experimental constraints on squark,
slepton, 
gluino, chargino, neutralino, and Higgs boson masses, on the $Z^0$ 
width and on the \bsg\ branching ratio
\cite{PDG,LEP2,CLEO}. The last row in Table~\ref{tab:scans}
gives the number of models which pass all experimental constraints.
We include the most recent LEP2 constraints \cite{LEP2} of
which the most important one is
\begin{equation}
  \label{eq:LEPmcha}
  m_{\chi^+} > 85 \mbox{ GeV} .
\end{equation}
This bound effectively excludes most of the higgsinos lighter than
the $W$ studied in Refs.~\cite{MizutaYamaguchi,NeuLoop1}. LEP2 also
puts a new constraint on the lightest Higgs boson mass,
\begin{equation} \label{eq:H2constraint}
  m_{H_2^0} > 62.5 \mbox{ GeV,}
\end{equation}
valid for all $\alpha$ and $\beta$. This constraint could be made
more stringent if allowed to depend on $\sin^2(\beta-\alpha)$, but
we do not include this more refined version because in this study we
are not very sensitive to this constraint.


\section{Results}
\label{sec:Results}

\begin{figure}
  \centerline{\epsfig{file=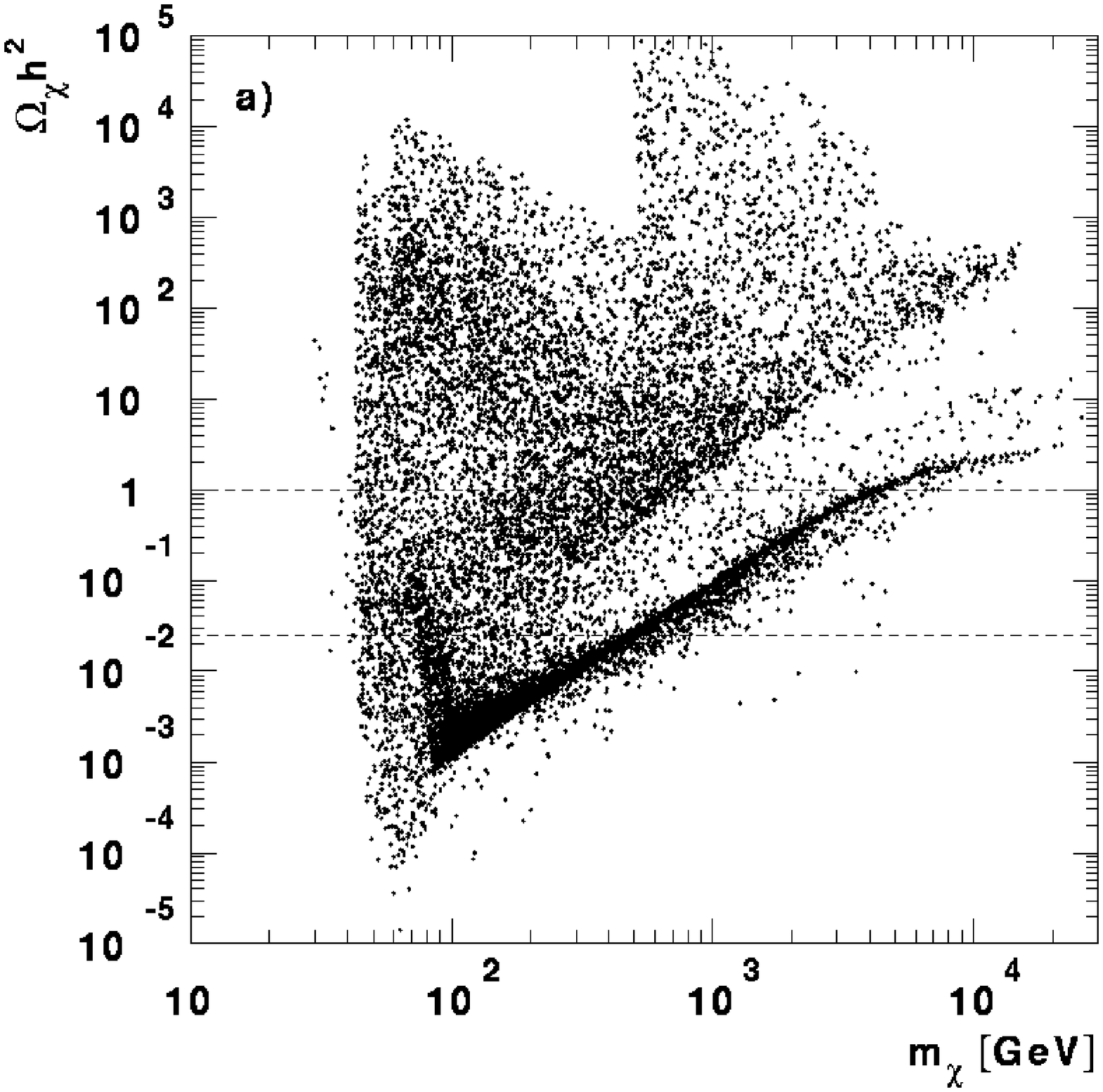,width=0.49\textwidth}
  \epsfig{file=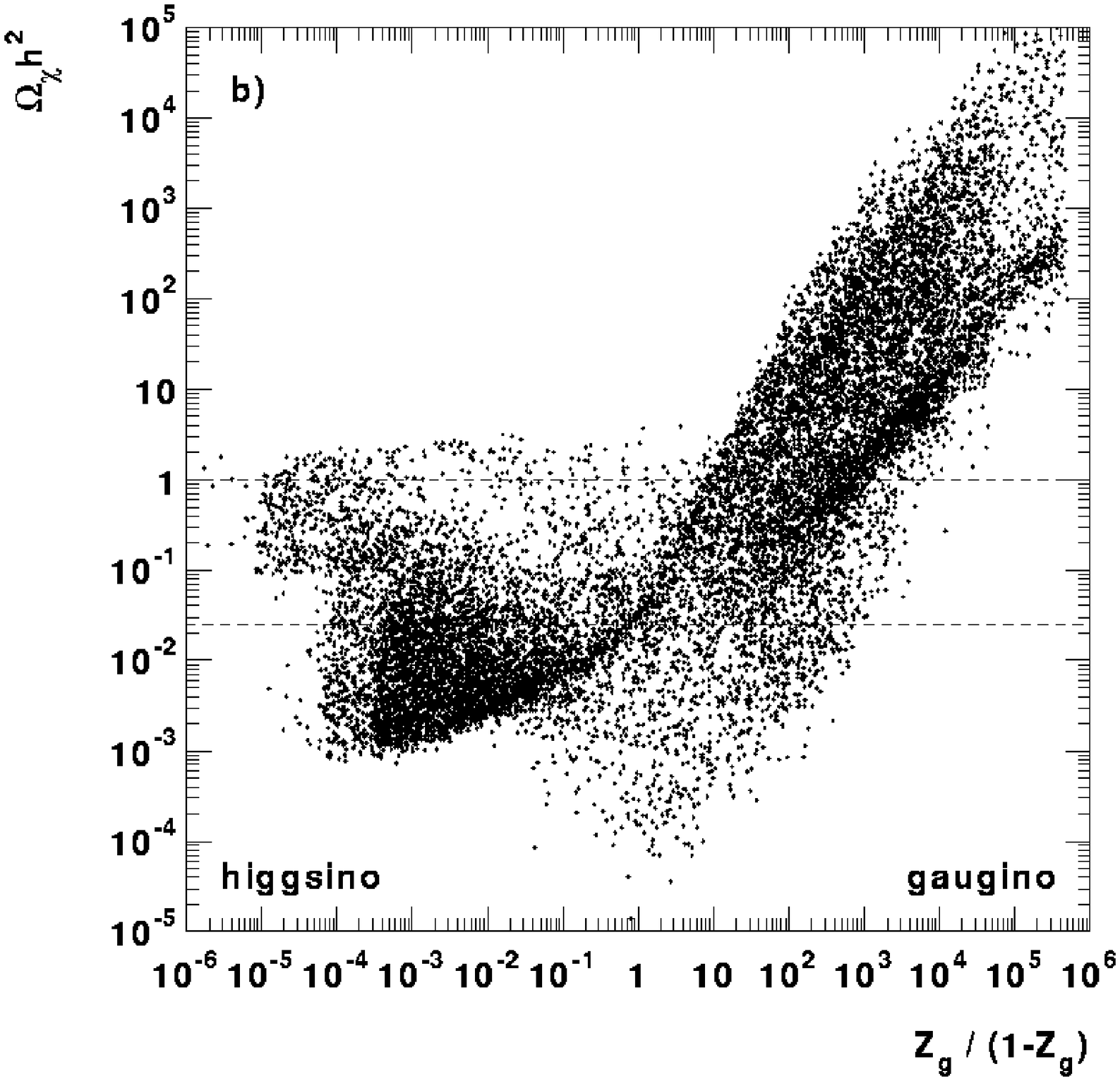,width=0.49\textwidth}}
  \caption{
    Neutralino relic density including neutralino and chargino
    coannihilations versus (a) neutralino mass $m_\chi$ and (b)
    neutralino composition $Z_g/(1-Z_g)$.  The horizontal lines
    bound the cosmologically interesting region $0.025 < \Omega_\chi
    h^2 <1$.}
  \label{fig:oh2vsmx}
\end{figure}

\begin{figure}
  \centerline{\epsfig{file=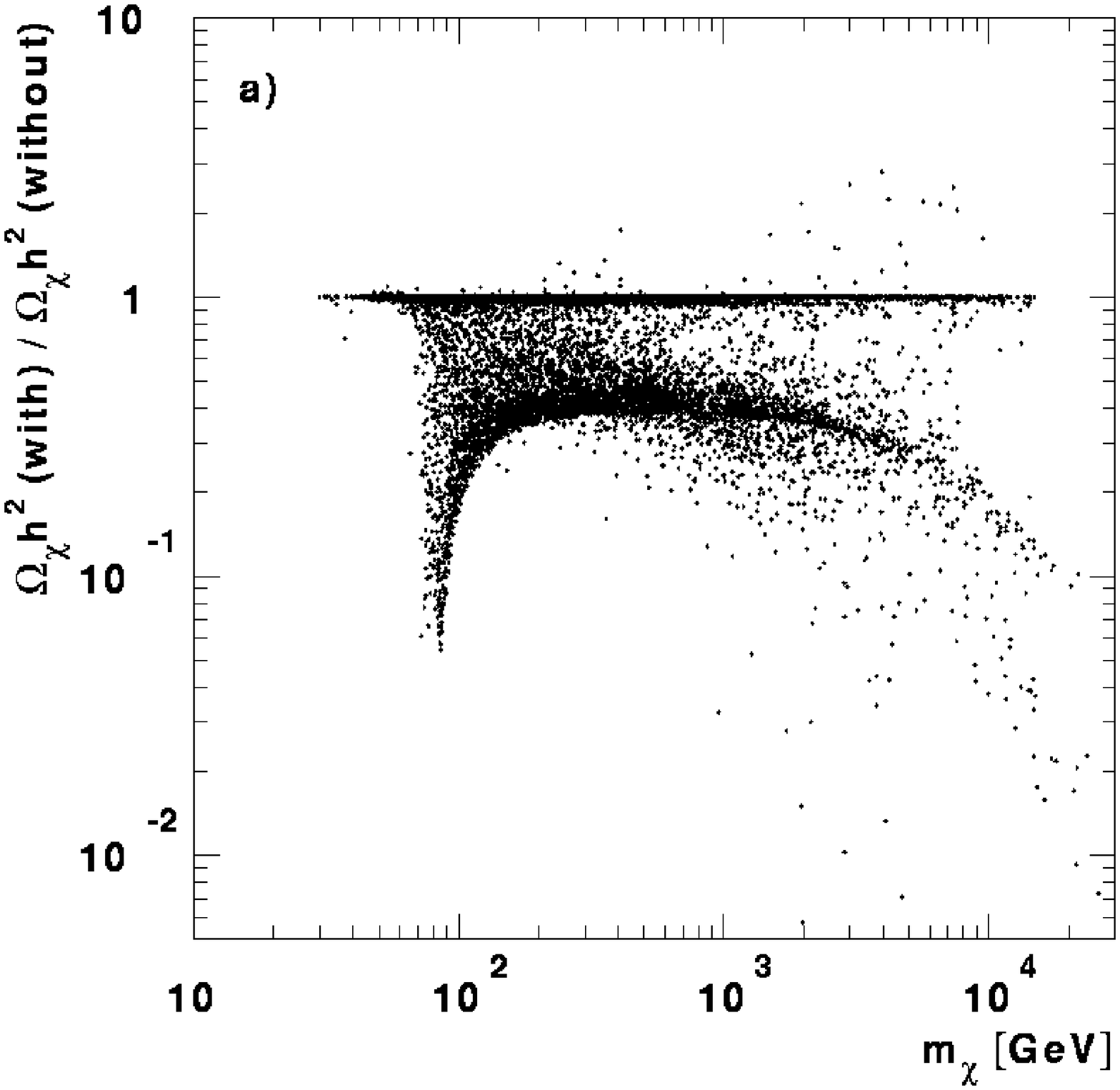,width=0.49\textwidth}
  \epsfig{file=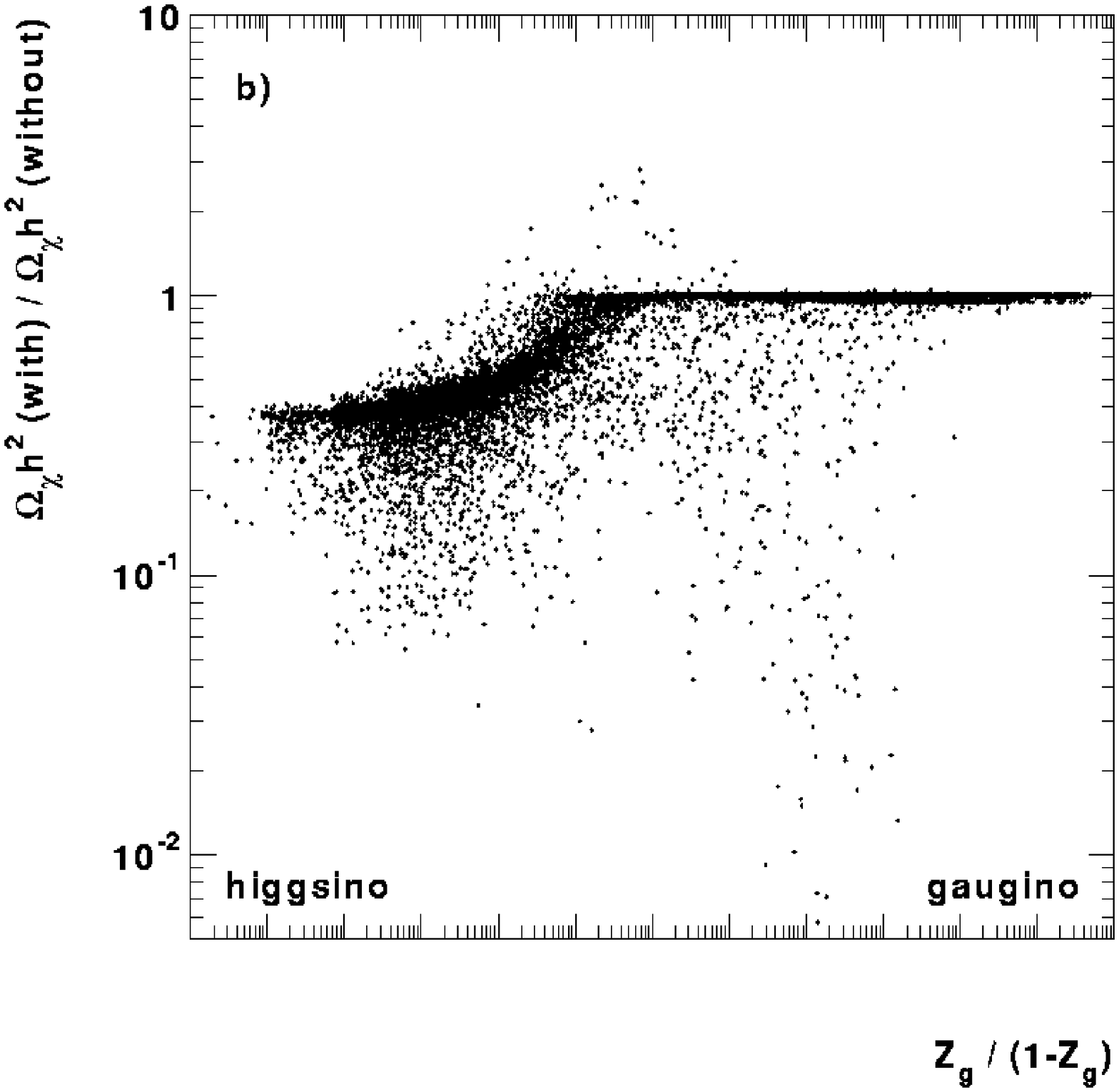,width=0.49\textwidth}}
  \caption{
    Ratio of the neutralino relic densities with and without
    neutralino and chargino coannihilations versus (a) neutralino
    mass $m_\chi$ and (b) neutralino composition $Z_g/(1-Z_g)$. }
  \label{fig:ratiovsmx}
\end{figure}

We now present the results of our relic density calculations for all
the models in Table~\ref{tab:scans}.  This is the first detailed
evaluation of the neutralino relic density including neutralino and
chargino coannihilations for general neutralino masses and
compositions. So we focus on the effect of coannihilations.

Fundamentally, we are interested in how the inclusion of
coannihilations modifies the cosmologically interesting region and the
cosmological bounds on the neutralino mass. We define the
cosmologically interesting region as $0.025 < \Omega_{\chi} h^2 < 1$.
In this range of $\Omega_\chi h^2$ the neutralino can constitute most
of the dark matter in galaxies and the age of the Universe is long
enough to be compatible with observations.  The lower bound of 0.025
is somewhat arbitrary, and even if $\Omega_{\chi}h^2$ would be less
than $0.025$ the neutralinos would still be relic particles, but only
a minor fraction of the dark matter in the Universe.

We start with a short general discussion and then present more details
in the following subsections.

Fig.~\ref{fig:oh2vsmx} shows the neutralino
relic density $\Omega_\chi h^2$ with coannihilations included versus
the neutralino mass $m_\chi$ and the neutralino composition $Z_g/(1-Z_g)$.
 The lower edge on neutralino masses comes essentially
from the LEP bound on the chargino mass, Eq.~(\ref{eq:LEPmcha}). The few
scattered points at the smallest masses have low $\tan\beta$. The
bands and holes in the point distributions, and the lower edge in
$Z_g/(1-Z_g)$, are mere artifacts of our
sampling in parameter space.

The neutralino is a good dark matter candidate in the region
limited by the two horizontal lines (the cosmologically interesting
region). There are clearly models with cosmologically interesting
relic densities for a wide range of neutralino masses (up to 7 TeV)
and compositions (up to $10^{-4}$ in higgsino fraction $Z_h=1-Z_g$).
A plot of the cosmologically interesting region in the neutralino
mass--composition plane is in subsection~\ref{sec:cosmregion}.

The effect of neutralino and chargino coannihilations on the value of
the relic density is summarized in Fig.~\ref{fig:ratiovsmx}, where we
plot the ratio of the neutralino relic densities with and without
coannihilations versus the neutralino mass $m_\chi$ and the
neutralino composition $Z_g/(1-Z_g)$. In many models, coannihilations
reduce the relic density by more than a factor of ten, and in some
others they increase it by a small factor. Coannihilations increase
the relic density if the effective annihilation cross section
$\langle \sigma_{\rm eff} v \rangle < \langle \sigma_{11} v_{11}
\rangle$. Recalling that $\langle \sigma_{\rm eff} v \rangle$ is the
average of the coannihilation cross sections (see
Eq.~(\ref{eq:sigmaveff})), this occurs when most of the
coannihilation cross sections are smaller than $\langle \sigma_{11}
v_{11} \rangle$ and the mass differences are small.

Table~\ref{tab:reprmod} lists some representative models where
coannihilations are important, one (or two) for each case described
in the following subsections, plus one model where coannihilations
are negligible. Example 1 contains a light higgsino-like neutralino,
example 2 a heavy higgsino-like neutralino.  Examples 3 and 4 have
$|\mu| \sim |M_1|$, and example 5 has a very pure gaugino-like
neutralino. Example 6 is a model with a gaugino-like neutralino for
which coannihilations are not important.

\begin{table}
\small
\begin{center}
\begin{tabular}{lrrrrrr} \hline
        & \multicolumn{1}{c}{light}  & 
         \multicolumn{1}{c}{heavy} & 
        \multicolumn{2}{c}{$|\mu|\sim|M_1|$} &
       \multicolumn{1}{c}{$|\mu| \gg |M_1|$} &
        \multicolumn{1}{c}{gaugino}  \\ 
        & \multicolumn{1}{c}{higgsino}  & 
         \multicolumn{1}{c}{higgsino} & 
        & &
       \multicolumn{1}{c}{bino} &
        \\ \hline
Example No.          & \multicolumn{1}{c}{1}
        & \multicolumn{1}{c}{2}      & \multicolumn{1}{c}{3} & 
      \multicolumn{1}{c}{4} & \multicolumn{1}{c}{5} & 
        \multicolumn{1}{c}{6}  \\ \hline
$\mu$ [GeV]          & $77.7$   & 1024.3 & 358.7     & 
414.7 & $-7776.7$    & $-1711.1$ \\
$M_2$ [GeV]          & $-441.4$ & 3894.1 & $-691.1$  &
$-1154.6$ & $133.5$ & 396.6 \\
$\tan \beta$         & 1.31     & 40.0   & 2.00      &
7.30   & 37.0       & 22.8 \\
$m_A$ [GeV]          & 656.8    & 737.2  & 577.7     &
828.9 & 2039.5     & 435.1 \\
$m_0$ [GeV]          & 610.8    & 1348.3 & 1080.9    &
2237.9 & 4698.0   & 2771.6 \\
$A_b/m_0$            & $-1.77$  & $-1.53$& $-1.03$   &
$-1.26$ & $0.46$    & 1.97 \\
$A_t/m_0$            &  2.75    & $-2.01$& $-2.77$   &
$-0.80$  & $0.11$    & 0.52 \\ \hline
$m_{\chi_1^0}$ [GeV] &  76.3    & 1020.8 & 340.2     &
407.8  & 67.2      & 199.5 \\
$Z_g$                &  0.00160 & 0.00155 & 0.651    &
0.0262   & 0.999968  & 0.99933 \\
$m_{\chi_2^0}$ [GeV] &  96.3    & 1026.4 & 364.5     &
418.2  & 133.5     & 396.0 \\
$m_{\chi_1^+}$ [GeV] &  89.2    & 1023.7 & 362.2     &
414.1  & 133.5     & 396.0 \\
$\Omega_\chi h^2$ (no coann.) & 0.178  & 0.130  & 0.158  &
0.00522   & $1.33\times 10^4$       & 0.418 \\
$\Omega_\chi h^2$    & 0.0299 & 0.0388 & 0.0528 &
0.00905   & $1.15\times 10^4$      & 0.418 \\ \hline
\end{tabular}
\end{center}
\caption{
  Some representative models for which coannihilations are important
  (examples 1--5) and one model (example 6) for which they are not.  We
  give the seven model parameters, the masses of the lightest
  neutralinos and of the lightest chargino, the gaugino fraction of
  the lightest neutralino and the relic densities with and without
  coannihilations.}
\label{tab:reprmod}
\end{table}

We have looked for a simple general criterion for when coannihilations
should be included, one that goes beyond the trivial statement of an
almost degeneracy in mass between the lightest neutralino and other
supersymmetric particles. We have only found few rules of thumb, each
with important exceptions. We give here the best two. 

\begin{figure}
  \centerline{\epsfig{file=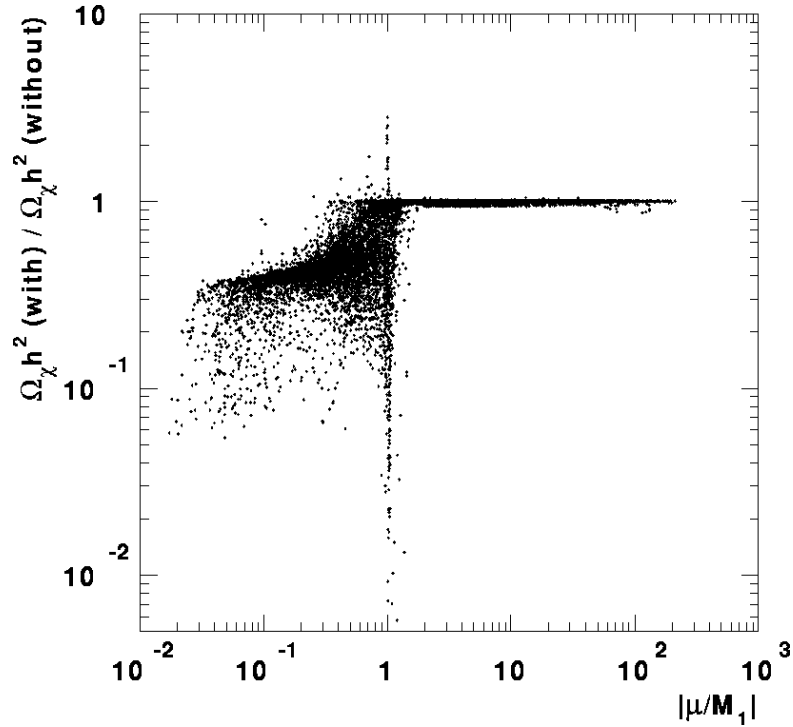,width=0.75\textwidth}} 
  \caption{ 
  Ratio of the relic densities with and without
  coannihilations versus $|\mu/M_1|$. Coannihilations are important 
  when $|\mu/M_1|\lsim 2$.}
  \label{fig:ratiovsmum1}
\end{figure}

The first rule of thumb is that when coannihilations are important,
$|\mu/M_1| \lsim 2$. But exceptions are found, as can be seen in
Fig.~\ref{fig:ratiovsmum1}, where we show the reduction in relic
density due to the inclusion of coannihilations as a function of
$|\mu/M_1|$. Notice that when $|\mu/M_1| \ll 1$, the neutralino is
higgsino-like; when $|\mu/M_1| \gg 1$, the neutralino is gaugino-like;
and when $|\mu/M_1| \sim 1$, the neutralino can be higgsino-like,
gaugino-like or mixed.

\begin{figure}
  \centerline{\epsfig{file=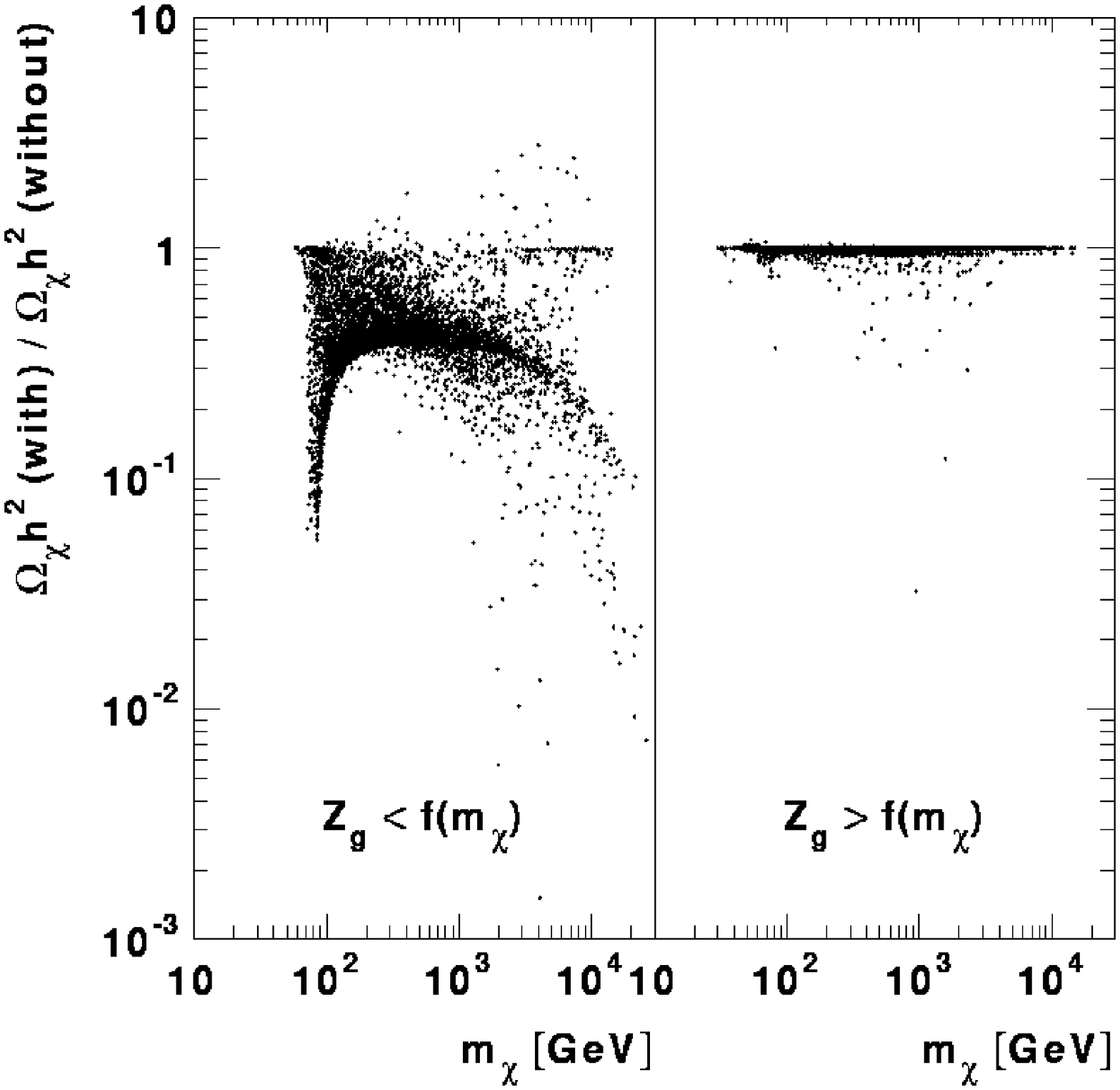,width=0.75\textwidth}} 
  \caption{ 
    Ratio of the relic densities with and without coannihilations
    versus neutralino mass $m_\chi$.  Coannihilations are generally
    not important when $Z_g > f(m_\chi)$, where $f(m_\chi)$ is the
    'second rule of thumb' given in the text.}
  \label{fig:ratiovsmx-zg}
\end{figure}

The second rule of thumb is that coannihilations are important when
$Z_g < 0.23$ for $m_\chi < 200$ GeV and when $Z_g/(1-Z_g) <
(m_\chi/300\mbox{ GeV})^{3}$ for $m_\chi > 200$ GeV. There are
exceptions to this rule, as can be seen in Fig.~\ref{fig:ratiovsmx-zg}
where the ratio of relic densities with and without coannihilations is
plotted versus the neutralino mass, the left panel for points
satisfying the present criterion, the right panel for those not
satisfying it.

In the following subsections, we present the cases where we found that
coannihilations are important and explain why. We first discuss
the already known case of light higgsino-like neutralinos, continue
with heavier higgsino-like neutralinos, the case $|\mu| \sim |M_1|$
and finally very pure gaugino-like neutralinos. We then end this
section by a discussion of the cosmologically interesting region.

\subsection{Light higgsino-like neutralinos}
\label{sec:lighthiggsino}

We first discuss light higgsino-like neutralinos, $m_\chi<m_W$, $Z_g <
0.01$, since coannihilation processes for these have been investigated
earlier by other authors \cite{MizutaYamaguchi, DreesNojiri,
  NeuLoop1}.

Mizuta and Yamaguchi~\cite{MizutaYamaguchi} stressed the great
importance of including coannihilations for higgsinos lighter than the
$W$ boson.  For these light higgsinos, neutralino-neutralino
annihilation into fermions is strongly suppressed whereas
chargino-neutralino and chargino-chargino annihilations into fermions
are not.  Since the masses of the lightest neutralino and
the lightest chargino are of the same order, the relic density is
greatly reduced when coannihilations are included. Mizuta and
Yamaguchi claim that because of this reduction light higgsinos are
cosmologically of no interest.

Drees and Nojiri~\cite{DreesNojiri} included coannihilations between
the lightest and next-to-lightest neutralino, but overlooked
those between the lightest neutralino and chargino, which are always
more important. In spite of this, they
concluded that the relic density of a higgsino-like neutralino will
always be uninterestingly small unless $m_\chi>500$ GeV or so.

Drees at al.~\cite{NeuLoop1} then re-investigated the relic density of
light higgsino-like neutralinos. They found that light higgsinos could
have relic densities as high as 0.2, and so be cosmologically
interesting, provided one-loop corrections to the neutralino masses
are included.

\begin{figure}
  \centerline{\epsfig{file=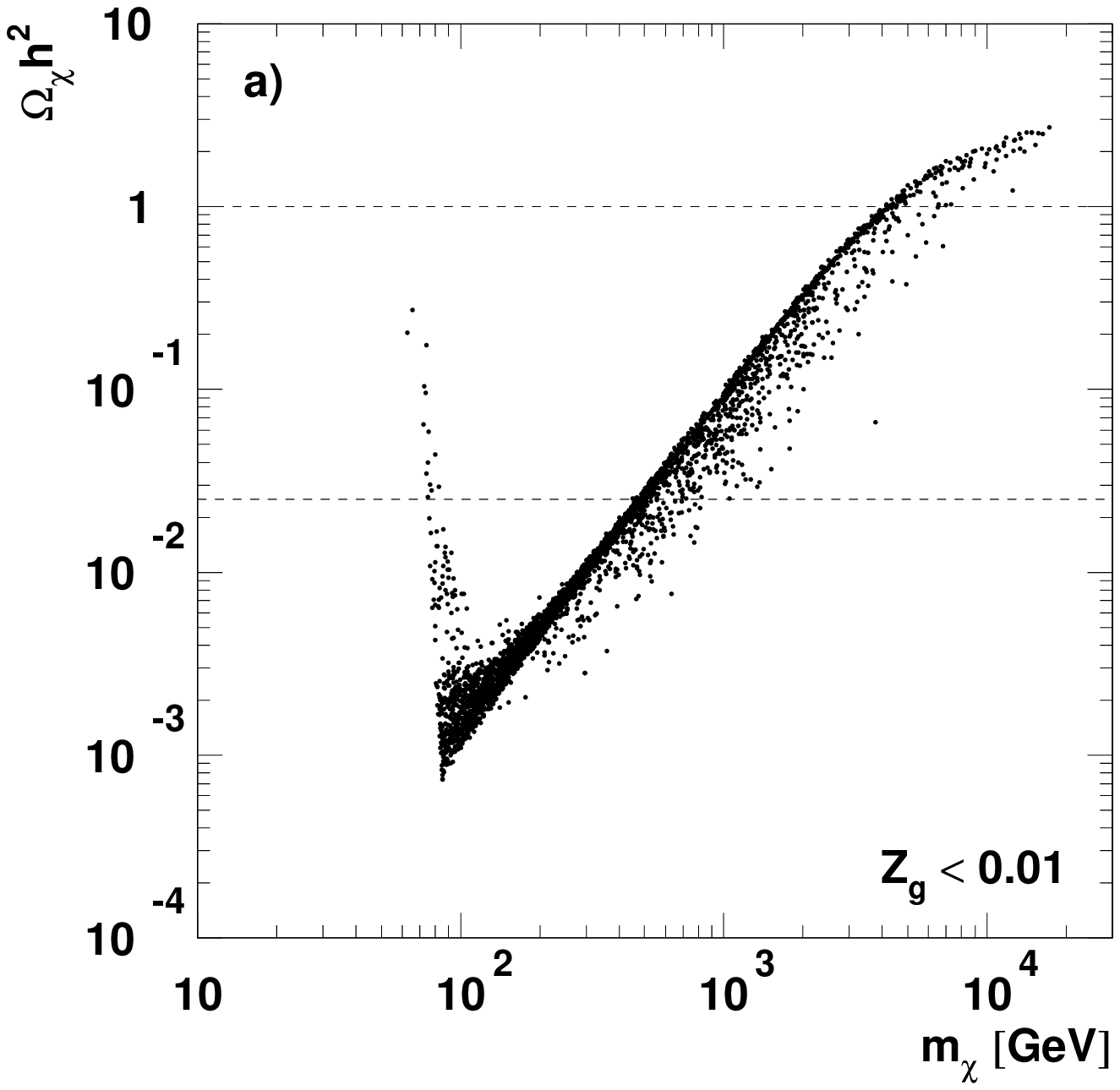,width=0.49\textwidth}
  \epsfig{file=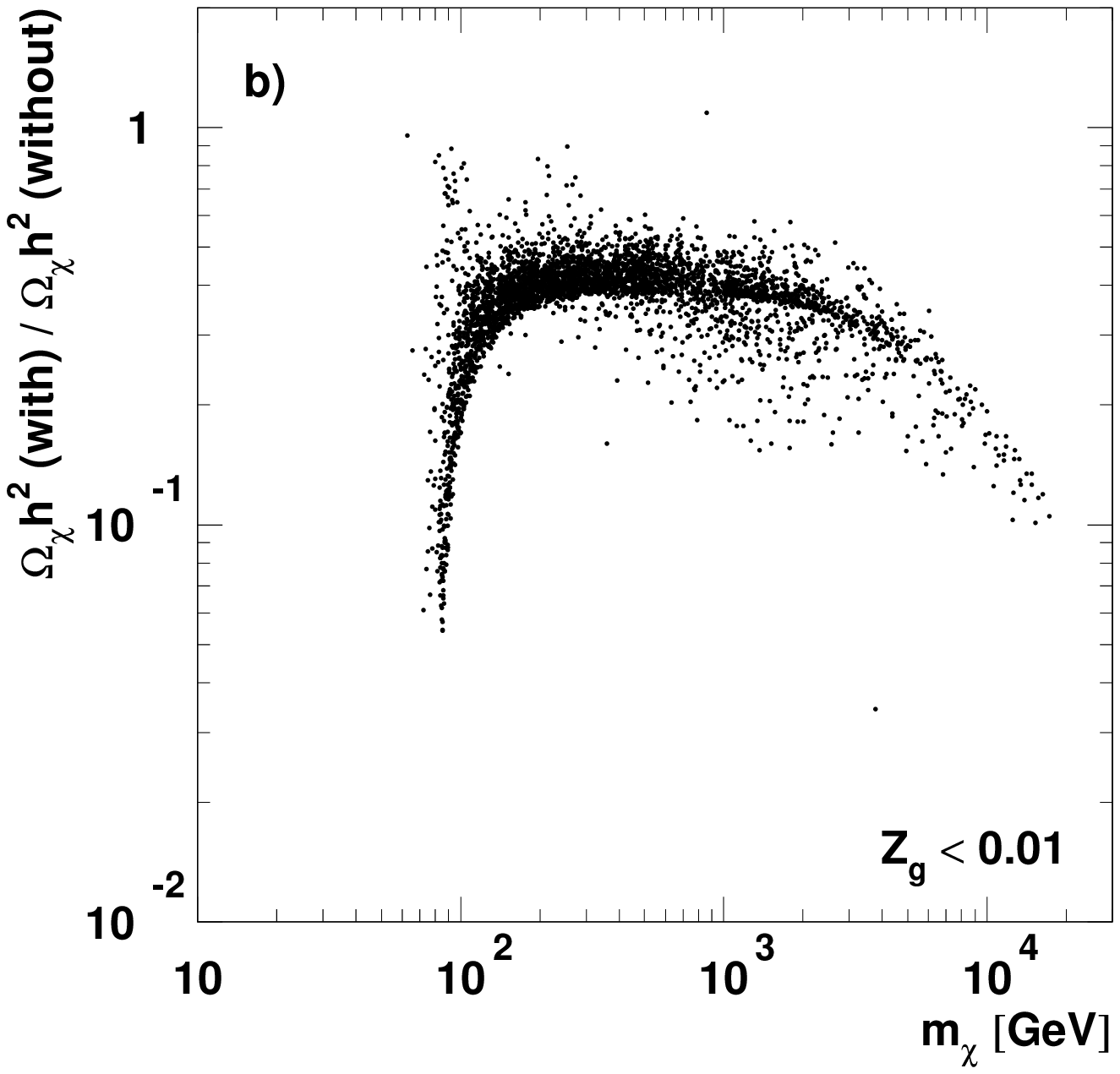,width=0.49\textwidth}}
  \caption{
    For higgsino-like neutralinos ($Z_g<0.01$) we show (a) the relic
    density with coannihilations included and (b) the ratio of the
    relic densities with and without coannihilations versus the
    neutralino mass.  The horizontal lines in (a) limit the
    cosmologically interesting region $0.025 < \Omega_\chi h^2 <1$.}
  \label{fig:oh2vsmxh}
\end{figure}

We agree with these papers qualitatively, but we reach different
conclusions. We show our results in Fig.~\ref{fig:oh2vsmxh}, where we
plot the relic density of higgsino-like neutralinos versus their
mass with coannihilations included, as well as the ratio between the
relic densities with and without coannihilations. The Mizuta and
Yamaguchi reduction can be seen in Fig.~\ref{fig:oh2vsmxh}b below 100
GeV, but due to the recent LEP2 bound on the chargino mass the effect
is not as dramatic as it was for them. If for the sake of comparison
we relax the LEP2 bound, the reduction continues down to $10^{-5}$ at
lower higgsino masses and we confirm qualitatively the Mizuta and
Yamaguchi conclusion --- coannihilations are very important for light
higgsinos --- but we differ from them quantitatively since we find
models in which light higgsinos have a cosmologically interesting
relic density.  For the specific light higgsino models in Drees et
al.~\cite{NeuLoop1} we agree on the relic density to within 20--30\%.
We find however other light higgsino-like models with higher
$\Omega_\chi h^2 \sim 0.3$, even without including the loop
corrections to the neutralino masses.

So there is a window of light higgsino models, $m_\chi \sim 75$ GeV,
that are cosmologically interesting. All these models have $\tan \beta
\lsim 1.6$ and those with the highest relic densities have $\tan \beta
\sim 1.2$.  These models escape the LEP2 bound on the chargino mass,
$m_{\chi^+} \sim 85$ GeV, because for $\tan \beta \lsim 2$ the mass of
the lightest neutralino can be lower than the mass of the lightest
chargino by tens of GeV\@.  By the same token, coannihilation
processes are not so important and the relic density in these models
remains cosmologically interesting.  Most of these models will be
probed in the near future when LEP2 runs at higher energies, but some
have too large a chargino mass ($m_\chi^+>95$ GeV) and too large an $H_2^0$
boson mass ($m_{H_2^0}>90$ GeV) to be tested at LEP2\@. Thus $\sim 75$
GeV higgsinos with $\tan\beta \lsim 2$ may remain good dark matter
candidates even after LEP2.

\subsection{Heavy higgsino-like neutralinos}
\label{sec:heavyhiggsino}

Coannihilations for higgsino-like neutralinos heavier than the W
boson have been mentioned by Drees and Nojiri~\cite{DreesNojiri}, who
argued that they should not change the relic density by much, and by
McDonald, Olive and Srednicki~\cite{McDonald}, who warn that they
might change it by an estimated factor of 2. We are the first to
present a quantitative evaluation in this case. We typically find a
decrease by factors of 2--5, and in some models even by a factor of
10 (see the right hand part of Fig.~\ref{fig:oh2vsmxh}b).

For $m_{\chi}>m_{W}$, the lightest and next-to-lightest neutralinos
and the lightest chargino are close in mass, and they annihilate into
$W$-bosons besides fermion pairs. While the annihilation and
coannihilation cross sections into $W$-pairs are comparable, the
coannihilation of $\chi_1^0 \chi_2^0$, $\chi_1^0 \chi^+_1$ and $\chi_2^0
\chi^+_1$ into fermion pairs is stronger than the $\chi_1^0 \chi_1^0\to
f\bar{f}$ annihilation. This gives the increase in the effective
annihilation rate that we observe.

As a result, the smallest and highest masses for which higgsino-like
neutralinos heavier than the $W$ boson are good dark matter candidates
shift up from 300 to 450 GeV and from 3 to 7 TeV respectively.

Together with the result in the previous subsection, we conclude that
higgsino-like neutralinos ($Z_g<0.01$) can be good dark matter
candidates for masses in the ranges 60--85 GeV and 450--7000 GeV.

\subsection{Models with $|\mu| \sim |M_1|$}
\label{sec:mum1}

\begin{figure}
  \centerline{\epsfig{file=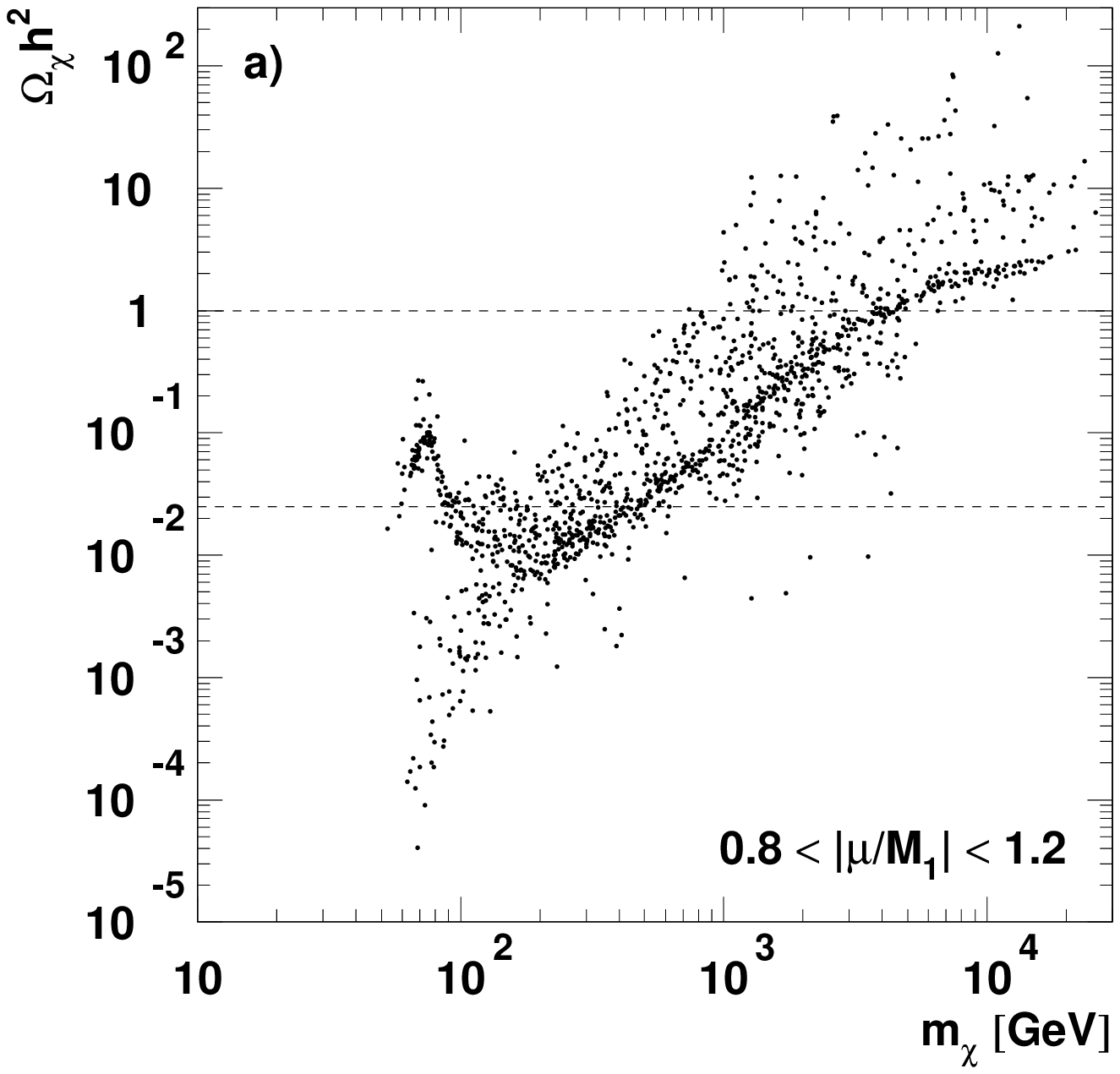,width=0.49\textwidth}
  \epsfig{file=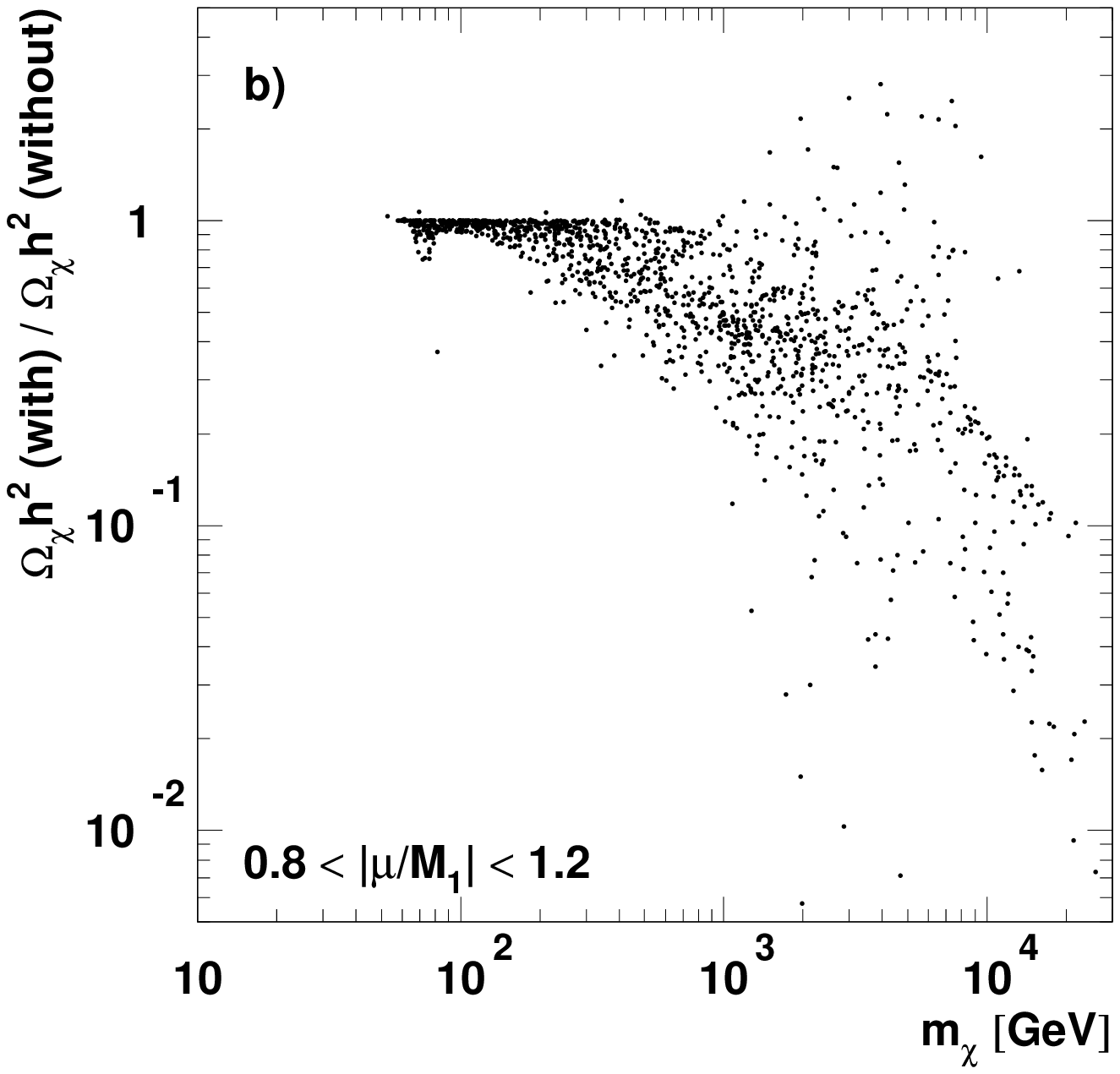,width=0.49\textwidth}}
  \caption{
    For neutralinos with $0.8<|\mu/M_1|<1.2$ we show (a) the relic
    density with coannihilations included and (b) the ratio of the
    relic densities with and without coannihilations versus the
    neutralino mass.  The horizontal lines in (a) limit the
    cosmologically interesting region $0.025 < \Omega_\chi h^2 <1$.}
  \label{fig:oh2vsmxmum1}
\end{figure}

Coannihilations for mixed or gaugino-like neutralinos have not been
included in earlier calculations. It has been believed that they are
not very important in these cases.  On the contrary, when $|\mu| \sim
|M_1|$ and $m_\chi \gsim m_W$, there is a very pronounced mass
degeneracy among the three lightest neutralinos and the lightest
chargino. The ensuing coannihilations can decrease the relic density
by up to two orders of magnitude or even \emph{increase} it by up to a
factor of 3.  This is easily seen in Fig.~\ref{fig:ratiovsmum1} as the
vertical strip at $|\mu/M_1| \sim 1$. In Fig.~\ref{fig:oh2vsmxmum1}
the relic density including coannihilations and the ratio of the relic
density with to that without coannihilations are shown versus the
neutralino mass for models with $0.8 < |\mu/M_1| < 1.2$.

We recall that in models with $|\mu|\sim |M_1$ the lightest neutralino
can be higgsino-like, mixed or gaugino-like.
If the lightest neutralino is mixed ($Z_g \sim 0.5$), coannihilations
can increase the relic density, whereas if it is more higgsino-like
or gaugino-like they will decrease it. This because the annihilation
cross section for mixed neutralinos is generally higher than those
for higgsino-like or gaugino-like neutralinos.

The largest decrease we see for this kind of models is when $|M_1|$
is slightly less than $|\mu|$ and both are in the TeV region. In
this case, the lightest neutralino is a very pure bino, and its
annihilation cross section is very suppressed since it couples
neither to the $Z$ nor to the $W$ boson. The chargino and other
neutralinos close in mass have much higher annihilation cross
sections, and thus coannihilations between them greatly reduce the
relic density. This big reduction suffices to lower $\Omega_\chi
h^2$ to cosmologically acceptable levels if $Z_g < 0.96$.  This
reduction does not occur for masses much lower than a TeV, because
the terms in the neutralino mass matrix proportional to the $W$ mass
prevent such pure bino states and such severe mass degeneracy.

To conclude, when $|\mu| \sim |M_1|$, coannihilations are very
important no matter if the neutralino is higgsino-like, mixed or
gaugino-like. The relic density can be cosmologically interesting for
these models as long as the gaugino fraction $Z_g<0.96$: these
neutralinos are good dark matter candidates.

\subsection{Gaugino-like neutralinos with $|\mu| \gg |M_1|$}
\label{sec:heavygaugino}

When $|\mu| \gg |M_1|$, the lightest neutralino is a very pure
gaugino.  According to the GUT relation Eq.~(\ref{eq:M1}), the
supersymmetric particles next in mass, the next-to-lightest neutralino
and the lightest chargino, are twice as heavy.  So we expect that
coannihilations between them are of no importance.\footnote{In
  models with non-universal gaugino masses, the lightest gaugino-like
  neutralino can be almost degenerate with the lightest chargino, and
  coannihilations can be important, as examined e.g. in
  Ref.~\cite{ChenDreesGunion}.} In fact, as discussed in
section~\ref{sec:NumMethods}, coannihilations would need to increase
the effective cross section by several orders of magnitude for these
large mass differences.

This actually happens in some cases, the small spread at $|\mu/M_1|
\simeq 130$ in Fig.~\ref{fig:ratiovsmum1}.  In these models, the
lightest neutralino is a very pure bino ($Z_g>0.999$) and the squarks
are heavy. Its annihilation to fermions is suppressed by the heavy
squark mass, and its annihilation to $Z$ and $W$ bosons is either
kinematically forbidden or extremely suppressed because a pure bino
does not couple to $Z$ and $W$ bosons.  On the other side, the
lightest chargino is a very pure wino, which annihilates to gauge
bosons and fermions very efficiently. The huge increase in the
effective cross section, compensated by the large mass difference,
reduces the relic density by 10--20\%.  However, the relic density
before introducing coannihilations was of the order of $10^3$--$10^4$,
and this small reduction is not enough to render these special cases
cosmologically interesting.

\subsection{Cosmologically interesting region}
\label{sec:cosmregion}

\begin{figure}
  \epsfig{file=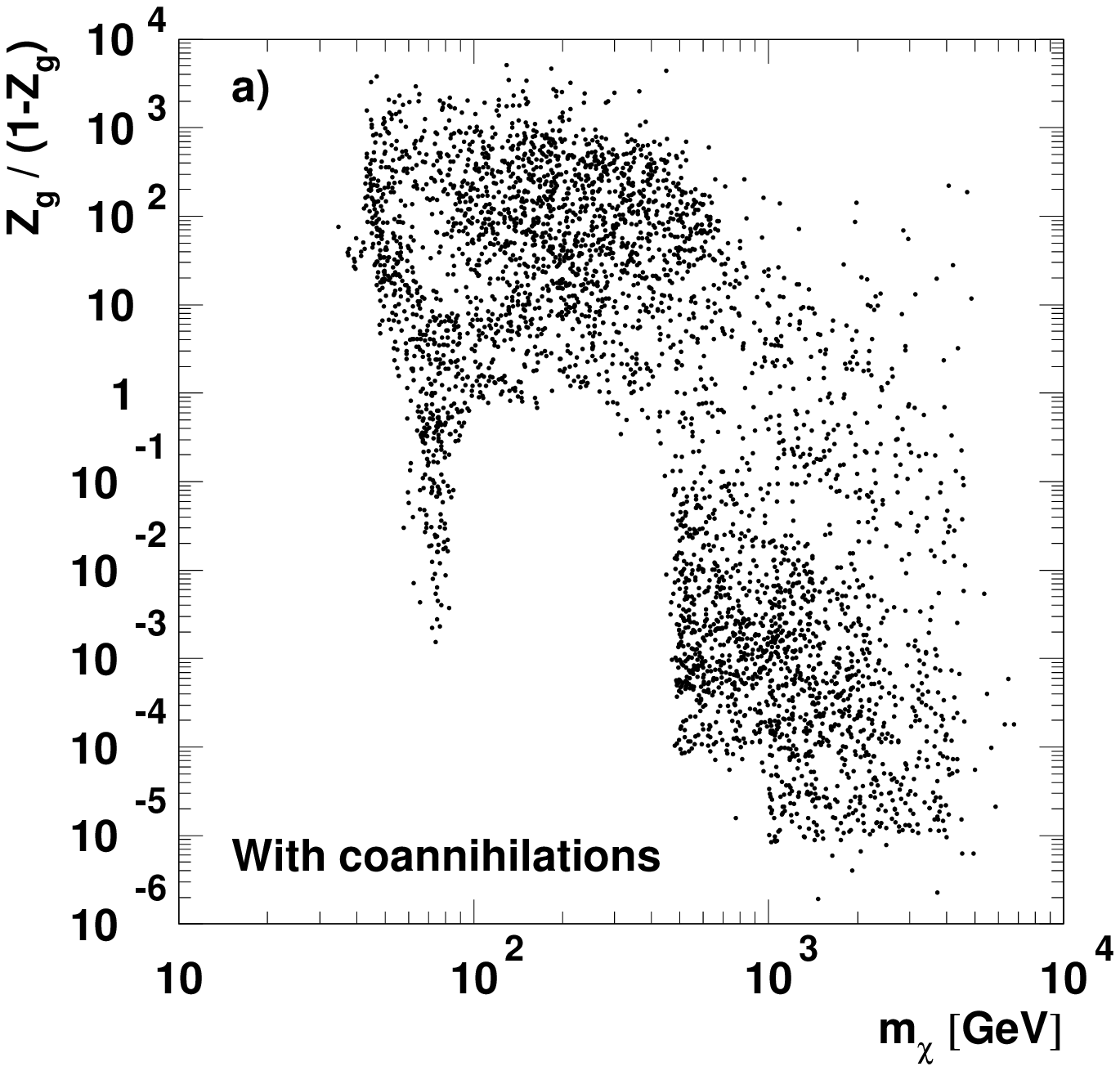,width=0.49\textwidth}
  \epsfig{file=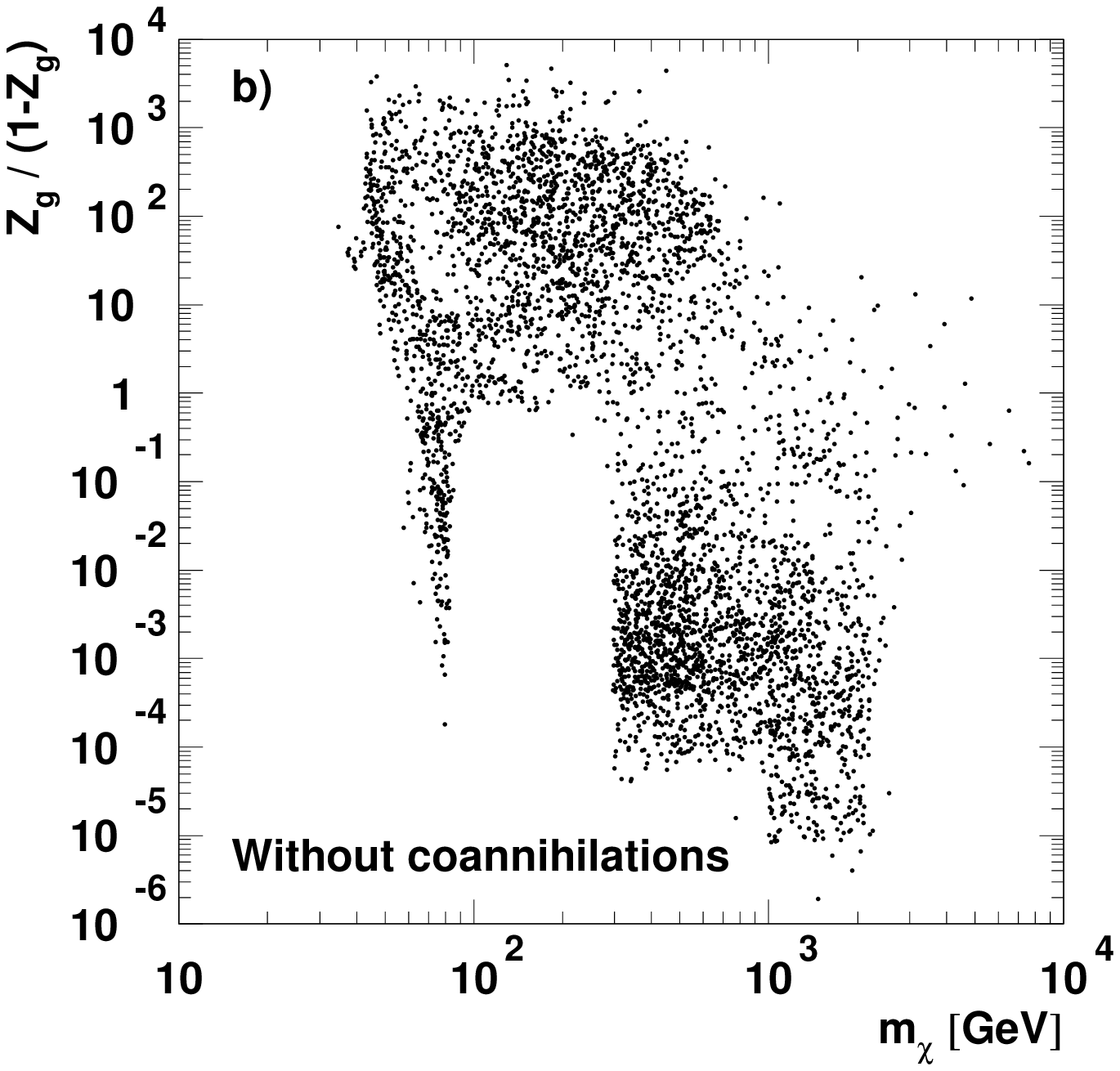,width=0.49\textwidth}
  \caption{Neutralino masses $m_\chi$ and compositions $Z_g/(1-Z_g)$
    for cosmologically interesting models (a) with and (b) without
    inclusion of coannihilations.}
  \label{fig:cosmregion}
\end{figure}

We now summarize when the neutralino is a good dark matter
candidate.  Fig.~\ref{fig:cosmregion} shows the cosmologically
interesting region $0.025 < \Omega_\chi h^2 <1$ in the neutralino
mass--composition plane $Z_g/(1-Z_g)$ versus\ $m_\chi$.

The light higgsino-like region does not extend to the left and down
because of the LEP2 bound on the chargino mass. The lower edge in 
gaugino fraction at $Z_{g} \sim 10^{-5}$ is the border of our survey
(how high $|M_{2}|$ is allowed to be). The upper limit on $Z_g$ and
the upper limit on the neutralino mass come from the requirement
$\Omega_\chi h^2<1$. The hole for higgsino-like neutralinos with
masses 85--450 GeV comes from the requirement $\Omega_\chi h^2>0.025$.

We see that coannihilations change the cosmologically interesting
region in the following aspects: the region of light higgsino-like
neutralinos is slightly reduced and the big region of heavier
higgsinos is shifted to higher masses, the lower boundary shifting
from 300 GeV to 450 GeV and the upper boundary from 3 TeV to 7 TeV.

The fuzzy edge at the highest masses is due to models in which the
squarks are close in mass to the lightest neutralino, in which case
$t$- and $u$-channel squark exchange enhances the annihilation cross
section. In these rather accidental cases, coannihilations with
squarks are expected to be important and enhance the effective cross
section even further. Thus, the upper bound of 7 TeV on the neutralino mass
may be an underestimate.

\section{Conclusions}
\label{sec:Concl}

We have performed a detailed evaluation of the relic density of the
lightest neutralino, including for the first time all two-body
coannihilation processes between neutralinos and charginos for
general neutralino masses and compositions. 

We have generalized the relativistic formalism of Gondolo and
Gelmini~\cite{GondoloGelmini} to properly treat (sub)threshold and
resonant annihilations also in presence of coannihilations. We have
found that coannihilations can formally be considered as thresholds
in a suitably defined Lorentz-invariant effective annihilation rate.

Our results confirm qualitatively the conclusion of Mizuta and
Yamaguchi~\cite{MizutaYamaguchi}: the inclusion of coannihilations
when $m_{\chi}<m_{W}$ is very important when the neutralino is
higgsino-like.  In contrast to their calculation we do however find a
window of cosmologically interesting higgsino-like neutralinos where
the masses are $m_{\chi} \sim 75$ GeV and $\tan \beta \lsim 1.6$\@.
This is due primarily to a milder mass degeneracy at low $\tan \beta$,
and secondarily to the one-loop corrections to the neutralino masses
pointed out in Ref.~\cite{NeuLoop1}.

We also find that coannihilations are important for heavy
higgsino-like neutralinos, $m_\chi>m_W$, for which the relic density
can decrease by typically a factor of 2--5, but sometimes even by a
factor of 10.  Higgsino-like neutralinos with $m_\chi>450$ GeV can
have $\Omega_{\chi} h^2>0.025$ and hence make up at least a major part
of the dark matter in galaxies.

When $|\mu| \sim |M_1|$, coannihilations will always be important:
they can decrease the relic density by up to a factor of 100 or even
increase it by up to a factor of 3. In these models, the neutralino is
either higgsino-like, mixed or gaugino-like, and when the gaugino
fraction $Z_g<0.96$, the relic density can be cosmologically
interesting.

Coannihilations between neutralinos and charginos increase the
cosmological upper limit on the neutralino mass from 3 to 7 TeV.
Coannihilations with squarks might increase it further.

Coannihilation processes must be included for a correct evaluation of
the neutralino relic density when $|\mu| \gg |M_1|$ and when
$|\mu|\lsim 2 |M_1| $.  In the first case, the neutralino is a very
pure gaugino and its relic density overcloses the Universe. In the
second case, the neutralino is either higgsino-like, mixed or
gaugino-like, and for each of these types there are many models where
it is a good dark matter candidate. To establish this, the inclusion
of coannihilations has been essential.

\section*{Acknowledgements}

J.~Edsj{\"o} would like to thank L.~Bergstr{\"o}m for enlightening
discussions and comments. P.~Gondolo is grateful to J.~Silk for the
friendly hospitality and encouraging support during his visit to the
Center for Particle Astrophysics, University of California, Berkeley,
where part of this work was completed. We also thank L.~Bergstr{\"o}m
for reading parts of the manuscript.



\begin{thebibliography}{99}
 
\bibitem{CMBdetexp}
C.~Bennett et al., MAP home page, http://map.gsfc.nasa.gov/;
M.~Bersanelli et al., ESA Report D/SCI(96)3, PLANCK home page,
http://astro.estec.esa.nl/SA-general/Projects/Cobras/cobras.html

\bibitem{CMBdettheory}
R.J.~Bond, R.~Crittenden, R.L.~Davis, G.~Efstathiou and P.J.~Steinhardt,
Phys.\ Rev.\ Lett. {\bf 72} (1994) 13;
G.~Jungman, M.~Kamionkowski, A.~Kosowsky and D.N.~Spergel, 
Phys.\ Rev.\ Lett. {\bf 76} (1996) 1007, Phys.\ Rev.\ {\bf D54} (1996)
1332;
J.R.~Bond, G.~Efstathiou and M.~Tegmark, astro-ph/9702100.

\bibitem{relcalc} Some selected references are: H.~Goldberg, Phys.\ 
  Rev.\ Lett.\ {\bf 50} (1983) 1419; L.M.~Krauss, Nucl.\ Phys.\ {\bf
    B227} (1983) 556; J.~Ellis et al., Nucl.\ Phys.\ {\bf B238}
  (1984) 453; K.~Griest, Phys.\ Rev.\ {\bf D38} (1988) 2357 [erratum
  ibid {\bf D39} (1989) 3802]; J.~Scherrer and M.S.~Turner, Phys.\ 
  Rev.\ {\bf D33} (1986) 1585 [erratum ibid {\bf D34} (1986) 3263];
  K.~Griest, M.~Kamionkowski and M.S.~Turner, Phys.\ Rev.\ {\bf D41}
  (1990) 3565; G.B.~Gelmini, P.~Gondolo, and E.~Roulet, Nucl.\ Phys.\ 
  {\bf B351} (1991) 623; A.~Bottino et al., Astropart.\ 
  Phys.\ {\bf 1} (1992) 61, ibid.\ {\bf 2} (1994) 67; 
  R.~Arnowitt and P.~Nath, Phys.\ Lett.\ {\bf B299} (1993) 58, 
  {\bf B307} (1993) 403(E), Phys.\ Rev.\ Lett.\ {\bf 70} (1993) 3696;
  H.~Baer and
  M.~Brhlik, Phys.\ Rev.\ {\bf D53} (1996) 597. For further
  references, see G.~Jungman, M.~Kamionkowski, and K.~Griest, Phys.\ 
  Rep.\ {\bf 267} (1996) 195.

\bibitem{SWO}
M.~Srednicki, R.~Watkins and K.A.~Olive, Nucl.\ Phys.\ {\bf B310} (1988) 
693.
 
\bibitem{GriestSeckel}
K.~Griest and D.~Seckel, Phys.\ Rev.\ {\bf D43} (1991) 3191.
 
\bibitem{McDonald} 
J.~McDonald, K.A.~Olive and M.~Srednicki, Phys.\ Lett.\ {\bf B283}
  (1992) 80;

\bibitem{MizutaYamaguchi}
S.~Mizuta and M.~Yamaguchi, Phys.\ Lett.\ {\bf B298} (1993) 120.

\bibitem{DreesNojiri}
M.~Drees and M.~Nojiri, Phys.\ Rev.\ {\bf D47} (1993) 376.

\bibitem{NeuLoop1}
M.~Drees, M.M.~Nojiri, D.P.~Roy and Y.~Yamada, hep-ph/9701219.

\bibitem{GondoloGelmini}
P.~Gondolo and G.~Gelmini, Nucl.\ Phys.\ {\bf B360} (1991) 145.
 
\bibitem{haberkane}
H.E. Haber and G.L. Kane, Phys. Rep. {\bf 117} (1985) 75;  
J.F. Gunion and H.E. Haber, Nucl. Phys. {\bf B272} (1986) 1 [Erratum-ibid. 
{\bf B402} (1993) 567].

\bibitem{carena}
M.~Carena, J.R.~Espinosa, M.~Quir{\'o}s and C.E.M.~Wagner, Phys.\
Lett.\ {\bfseries B355} (1995) 209.

\bibitem{effpot}
M.~Berger, Phys.\ Rev.\ {\bf D41} (1990) 225; H.E.~Haber and R.~Hempfling,
Phys.\ Rev.\ Lett.\ {\bf 66} (1991) 1815; P.H.~Chankowski, S.~Pokorski and
J.~Rosiek, Phys.\ Lett.\ {\bf B281} (1992) 100; J.~Ellis, G.~Ridolfi and 
F.~Zwirner, Phys.\ Lett.\ {\bf B257} (1991) 83; Y.~Okada, M.~Yamaguchi and
T.~Yanagida, Phys.\ Lett.\ {\bf B262} (1991)  54;
A.~Brignole, Phys.\ Lett.\ {\bf B281} (1992) 284;
V.~Barger, M.S.~Berger and P.~Ohmann, Phys.\ Rev.\ {\bf D49} (1994) 4908;
J.~Kodaira, Y.~Yasui and K.~Sasaki, Phys.\ Rev.\ {\bf D50} (1994) 7035.

\bibitem{NeuLoop2}
D.~Pierce and A.~Papadopoulos, Phys.\ Rev.\ {\bf D50} (1994) 565, 
Nucl.\ Phys.\ {\bf B430} (1994) 278; A.B.~Lahanas, K.~Tamvakis and 
N.D.~Tracas, Phys.\ Lett.\ {\bf B324} (1994) 387.

\bibitem{B-functions}
M.~Drees, K.~Hagiwara and A.~Yamada, Phys.\ Rev.\ {\bf D45} (1992) 
1725.

\bibitem{PDG}
R.M.~Barnett et al. (Particle Data Group), Phys.\ Rev.\ {\bf D54} (1996) 1.
  
\bibitem{reduce}
{\sc Reduce} 3.5. A.C.~Hearn, RAND, 1993.

\bibitem{paolo} 
P.~Gondolo, in preparation.

\bibitem{gondoloedsjo}
P.~Gondolo and J.~Edsj\"o, in preparation.

\bibitem{LEP2} G.~Cowan (ALEPH Collaboration), talk presented at the
  special CERN particle physics seminar on physics results from the
  LEP run at 172 GeV, 25 February, 1997,
  http://alephwww.cern.ch/ALPUB/seminar/Cowan-172-jam/cowan.html

\bibitem{CLEO}
M.S.~Alam, et al.\ (CLEO Collaboration), Phys.\ Rev.\ Lett.\
{\bf 71} (1993) 674;  Phys.\ Rev.\ Lett.\ {\bf 74} (1995) 2885.

\bibitem{ChenDreesGunion}
C.H.~Chen, M.~Drees, and J.F.~Gunion, Phys.\ Rev.\ {\bf D55} (1997) 330.

\end{thebibliography}
\end{document}